\documentclass[letterpaper,twocolumn,appendixfloats]{emulateapj}

\newcommand{\nd}{\multicolumn{1}{c}{$\dots$}}
\newcommand{\ndr}{\multicolumn{1}{r}{$\dots$}}
\slugcomment{Submitted to ApJS on 29 July 2011; revised on 16 December 2011}

\shorttitle{2MASS Redshift Survey}
\shortauthors{Huchra {\it et al.}}

\begin{document}

\title{The 2MASS Redshift Survey - Description and Data Release}
\author{
John P. Huchra\altaffilmark{1,2},
Lucas M. Macri\altaffilmark{3},
Karen L. Masters\altaffilmark{4,5},
Thomas H. Jarrett\altaffilmark{6},\\
Perry Berlind\altaffilmark{2},
Michael Calkins\altaffilmark{2},
Aidan C. Crook\altaffilmark{7},
Roc Cutri\altaffilmark{5},
Pirin Erdo\u{g}du\altaffilmark{8},\\
Emilio Falco\altaffilmark{2},
Teddy George\altaffilmark{9},
Conrad M. Hutcheson\altaffilmark{10},
Ofer Lahav\altaffilmark{8},
Jeff Mader\altaffilmark{11},
Jessica D. Mink\altaffilmark{2},\\
Nathalie Martimbeau\altaffilmark{12},
Stephen Schneider\altaffilmark{13},
Michael Skrutskie\altaffilmark{14},
Susan Tokarz\altaffilmark{2} \&
Michael Westover\altaffilmark{15}
}
\altaffiltext{1}{This paper is mostly based on text written by John Huchra before his death in October 2010.}
\altaffiltext{2}{Harvard-Smithsonian Center for Astrophysics, 60 Garden Street, Cambridge, MA 02138, USA}
\altaffiltext{3}{George P. and Cynthia Woods Mitchell Institute for Fundamental Physics and Astronomy, Department of Physics and Astronomy, Texas A\&M University, 4242 TAMU, College Station, TX 77843, USA. lmacri@tamu.edu}
\altaffiltext{4}{Institute for Cosmology and Gravitation, University of Portsmouth, Dennis Sciama Building, Burnaby Road, Portsmouth, PO1 3FX, UK. karen.masters@port.ac.uk}
\altaffiltext{5}{SEPNet (South East Physics Network), United Kingdom}
\altaffiltext{6}{Infrared Processing and Analysis Center, California Inst. of Technology, 770 S Wilson Ave, Pasadena, CA 91125, USA}
\altaffiltext{7}{Microsoft Corp., 1 Microsoft Way, Redmond, WA 98052, USA}
\altaffiltext{8}{Department of Physics and Astronomy, University College London, London WC1E 6BT, United Kingdom}
\altaffiltext{9}{Canada-France-Hawaii Telescope, 65-1238 Mamalahoa Hwy, Kamuela, HI 96743, USA}
\altaffiltext{10}{Kavli Institute for Particle Astrophysics and Cosmology, Stanford University, Stanford, CA 94309, USA}
\altaffiltext{11}{Keck Observatory, 65-1120 Mamalahoa Hwy, Kamuela, HI 96743, USA}
\altaffiltext{12}{Plan\'etarium de Montr\'eal, 1000 rue Saint-Jacques, Montr\'eal, Qu\'ebec H3C 1G7, Canada}
\altaffiltext{13}{Department of Astronomy, University of Massachusetts, Amherst, MA 01003, USA}
\altaffiltext{14}{Department of Astronomy, University of Virginia, Charlottesville, VA 22904, USA}
\altaffiltext{15}{McKinsey \& Co., 1420 Fifth Ave, Ste 3100, Seattle, WA, 98101, USA}

\begin{abstract}
We present the results of the 2MASS Redshift Survey (2MRS), a ten-year project to map the full three-dimensional distribution of galaxies in the nearby Universe. The 2 Micron All-Sky Survey (2MASS) was completed in 2003 and its final data products, including an extended source catalog (XSC), are available on-line. The 2MASS XSC contains nearly a million galaxies with K$_s\leq 13.5$~mag and is essentially complete and mostly unaffected by interstellar extinction and stellar confusion down to a galactic latitude of $|b|=5^{\circ}$ for bright galaxies. Near-infrared wavelengths are sensitive to the old stellar populations that dominate galaxy masses, making 2MASS an excellent starting point to study the distribution of matter in the nearby Universe.

We selected a sample of 44,599 2MASS galaxies with K$_s\leq11.75$~mag and $|b|\geq5^{\circ}$ ($\geq 8^{\circ}$ towards the Galactic bulge) as the input catalog for our survey. We obtained spectroscopic observations for 11,000~galaxies and used previously-obtained velocities for the remainder of the sample to generate a redshift catalog that is 97.6\% complete to well-defined limits and covers 91\% of the sky. This provides an unprecedented census of galaxy (baryonic mass) concentrations within 300~Mpc.

Earlier versions of our survey have been used in a number of publications that have studied the bulk motion of the Local Group, mapped the density and peculiar velocity fields out to 50~$h^{-1}$~Mpc, detected galaxy groups, and estimated the values of several cosmological parameters.

Additionally, we present morphological types for a nearly-complete sub-sample of 20,860 galaxies with K$_s\leq11.25$~mag and $|b|\geq10^{\circ}$.
\end{abstract}


\section {Introduction}
Between the mid-1970s and the early 1980s, several discoveries were made based on innovations in detector technology and better understanding of galaxies that substantially changed our view of the nearby Universe. The cosmic microwave background (CMB) dipole was convincingly measured \citep{corey76,smoot77,cheng79}, the first large redshift surveys were started \citep[c.f.][]{davis82} and Virgo Infall was both convincingly predicted and measured \citep{devaucouleurs56,silk74,peebles76,aaronson82}. The kinematics of the Local Universe became a cosmological test and tool and -- with the realization that the Virgo supercluster was insufficient to explain the CMB dipole -- the search for the source of the flow (astronomy's Nile!) became a major cosmological quest.

In the 1980s this quest led to the discovery of even larger mass concentrations such as the Great Attractor \citep{burstein86,lyndenbell88} and the Shapley Supercluster \citep{tully84,tammann85} and the initiation of several very large scale redshift surveys based on IR and optical catalogs \citep[e.g.][]{strauss92,santiago95,saunders00}. Perforce then followed advanced distance surveys and catalogs \citep{mould93,willick97}. Sophisticated techniques were developed to analyze these surveys \citep{dekel90,zaroubi95} but despite reasonable data and thorough analyses, the source of the CMB dipole was not convincingly identified and there remained very significant conflicts between the results of different surveys \citep[e.g.,][]{schmoldt99}.

\begin{deluxetable*}{lcccrr}
\tabletypesize{\scriptsize}
\tablewidth{0pc} 
\tablecaption{Large Redshift Surveys of the Nearby Universe to date \label{tb:zsurveys}}
\tablehead{\colhead{Survey} &\colhead{Sky coverage} &\colhead{Depth$^a$} & \colhead{Selection}   & \colhead{\# gals.} & \colhead {Reference}\\ 
                            &\colhead{\% $4\pi$ sr}    &\colhead{($z$)} & \colhead{(band, flux)}& \colhead{($\times 10^3$)}  & \colhead{}} 
\startdata 
 CfA1       & 30\%      & 0.03 & B=14.5~mag      &     2.4 & \citet{delapparent86}\\
 ORS        & 60\%      & 0.03 & B=14.0~mag      &     8.5 & \citet{santiago95}   \\
 SSRS2+     & 60\%      & 0.04 & B=15.5~mag      &    23.6 & \citet{dacosta98} \& \\
 CfA2       &           &      &                 &         & \citet{huchra99}     \\
\hline
 IRAS PSCz  & 85\%      & 0.08 & 60$\mu$m=0.6~Jy &    16.1 & \citet{saunders00}   \\
 LCRS       & ~1\%      & 0.17 & R=17.5~mag      &    25.3 & \citet{shectman96}   \\
 2dF        & ~8\%      & 0.19 & b$_{\rm J}$=19.5~mag &   245.6 & \citet{colless01}    \\
 SDSS$^b$   & 35\%      & 0.33 & r=17.5~mag      &   943.6 & \citet{aihara11}     \\
\hline
 6dFGS      & 40\%      & 0.10 & K$_s$=12.65~mag &   124.6 & \citet{jones04,jones05,jones09} \\
 2MRS11.25  & 83\%      & 0.04 & K$_s$=11.25~mag &    20.6 & \citet{huchra05}    \\
 2MRS       & 91\%      & 0.05 & K$_s$=11.75~mag &    43.5 & {\bf this work}     
\enddata
\tablecomments{(a): 90\%-ile redshift value in catalog. (b): DR8 main galaxy sample.}
\end{deluxetable*}

Near the end of the 1990s, there remained a conflict between $\Omega_M$ on all measured scales and the $\Omega_M=1$ strongly predicted from inflation and cold dark matter models. Was the discrepancy real or were there problems with the data and/or the theory? Most of the community realized that {\it all} extant maps were tremendously biased, either by extinction or by wavelength (read ``young star formation,'' which dominates {\it both} blue and far-infrared light).  This was the explanation advocated by the theorists --- the galaxies being measured were not really tracing the mass. 

Fortunately, the overall $\Omega$ problem was solved soon thereafter with the discovery of dark energy \citep{riess98,perlmutter99} coupled with the accurate determination of the Hubble constant \citep{freedman01} and the measurement of the large scale geometry of the Universe through observations of fluctuations in the CMB \citep{spergel03}. Still, several very significant questions remain. Can we accurately (to a few percent) observationally account for the matter density in the nearby Universe? How is matter distributed? In particular, can we explain gravitationally the motion of the Milky Way with respect to the CMB? Do we understand the differences, if any, in the distribution of ordinary baryonic matter and dark matter (i.e., the bias function)? These questions are yet unanswered and clearly drive the detailed understanding of galaxy and large-scale structure formation and evolution. 

Despite all of the aforementioned work, even the galaxy density field of the Local Supercluster (LSC) is not in good shape. Despite high-quality data on the flow field, \citet{tonry00} found there are many missing elements to the model of the LSC, including possible local sources of the observed quadrupole field and the ``Local Anomaly.''

\section{The Two Micron All-Sky Survey}

The Two Micron All-Sky Survey \citep[2MASS,][]{skrutskie06} had its origins in a proposal to NASA for a ``Near InfraRed Astronomical Satellite'' by G.~Fazio, J.~Huchra, J.~Mould and collaborators in 1988.  The survey was eventually carried out by a team led by astronomers at the University of Massachusetts (UMass) using twin 1.3-m telescopes located at Mount Hopkins, Arizona (starting in 1997) and Cerro Tololo, Chile (starting in 1998). Scans were completed by 2001 and the final data release was made available in 2003 through IPAC\footnote{\url{http://www.ipac.caltech.edu/2mass/}}.

2MASS mapped the entire sky in the J, H, and K$_s$ bands, avoiding many of the observational biases that affected previous optical and far-infrared all-sky surveys. The effects of interstellar extinction are reduced by $10\times$ relative to the B-band and the spectral energy distributions of most galaxies peak at near-infrared wavelengths. Moreover, K-band luminosities are a useful proxy for baryonic mass as the stellar mass-to-light ratio is fairly constant across galaxy types at this wavelength \citep[e.g., within a factor of 2;][]{bell01}. This makes the near-infrared the spectral region of choice to map the distribution of matter in the nearby Universe.

The 2MASS photometric pipeline produced a complete and reliable extended-source catalog \citep[XSC,][]{jarrett00,jarrett04} of $\sim10^6$ objects with K$_s\leq13.5$~mag and a mean photometric accuracy better than 0.1~mag. Moreover, the database included information on the photometric structure of the galaxies (photometric profiles, axis ratios, etc.). 2MASS provided the first modern, all-sky, highly accurate catalog of galaxies. A few years later, the Sloan Digital Sky Survey \citep{york00} started to provide overlapping deeper optical data which eventually covered $\sim35$\% of the sky \citep{aihara11}, but 2MASS remains the only modern, survey which can be used to construct a uniform, all-sky, three-dimensional map of the local Universe.

Two decades before 2MASS, the first flux-limited all-sky galaxy catalog was created from observations by the IRAS satellite at $60\mu$m \citep{strauss90}. Since galaxies were unresolved by IRAS, the point source catalog formed the basis of a redshift survey \citep[PSCz][]{fisher95,saunders00}. Among other problems, the PSCz catalog gave little weight to ellipticals (which are dim at $60\mu$m because this wavelength is dominated by dusty star formation) and suffered from severe confusion in regions of high density. However, the uniform full-sky coverage was unique at the time.

\subsection{The Zone of Avoidance}

\begin{figure}[t]
\begin{center}
\includegraphics[angle=0,width=0.45\textwidth]{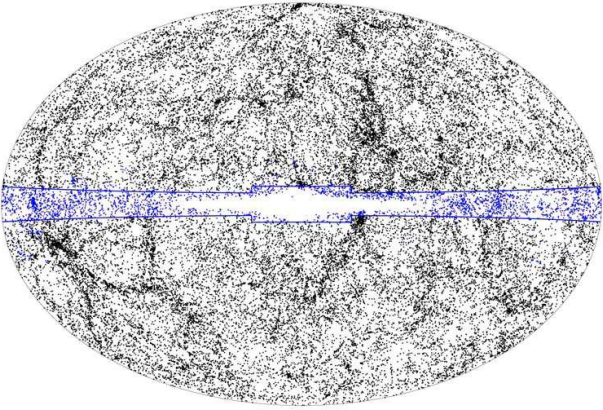}
\end{center}
\caption{Distribution of 2MASS galaxies with K$_s\leq11.75$~mag in Galactic coordinates (Aitoff projection). Blue dots represent galaxies outside our survey area.  Note that due to stellar confusion we cannot cover, even to this bright magnitude limit, the very central region of the galaxy but we do cover $\sim$91\% of the sky. \\ \label{fig:zoa}}
\end{figure}

\begin{figure}[t]
\hfill\includegraphics[angle=270,width=0.45\textwidth]{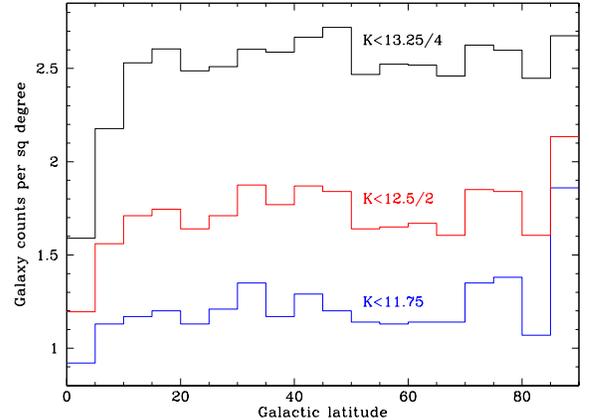}\hfill
\vspace{-2pt}
\caption{Surface number density vs galactic latitude for three cuts in the 2MASS XSC at K$_s=$11.75, 12.5 and 13.25~mag.  While the number counts drop sharply in the $5^\circ \leq |b| \leq 10^\circ$ bin for the 13.25~mag sample, the incompleteness is reduced for the 12.5~mag sample and it is essentially zero for the 11.75~mag sample. The upturn in all samples at 90$^\circ$ is due to the Coma supercluster.\label{fig:gcounts}}
\vspace{6pt}
\end{figure}

2MASS is an excellent probe of the zone-of-avoidance for bright galaxies, as was discussed in depth by \citet{huchra05}. Figure~\ref{fig:zoa} is an update of Fig.~8 from \citet{huchra05} showing the 2MASS XSC coverage at K$_s\leq11.75$~mag, limited only by confusion near the galactic center. Figure~\ref{fig:gcounts} is an update of Fig.~7 from \citet{huchra05} and shows the galaxy surface density versus galactic latitude for several magnitude limits.  At the bright magnitudes surveyed by 2MRS, the catalog is essentially complete to very low latitudes.

\section{The 2MASS Redshift Survey}

The primary extragalactic goal of 2MASS was to feed the next generation of all-sky redshift surveys to fully map the nearby Universe. To this end, we started a program in September 1997 to obtain the required spectroscopic data for a magnitude-limited sample of galaxies: the 2MASS Redshift Survey (2MRS). Our initial survey limits of K$_s=11.25$~mag and $|b|=10^\circ$ (20,860 galaxies; hereafter 2MRS11.25) were progressively increased to final values of K$_s=11.75$~mag and $|b|=5-8^\circ$ (44,599 galaxies; the full 2MRS), allowing us to steadily complete our view of the local universe. 

2MRS builds and improves on the previous generation of local surveys (see Table~\ref{tb:zsurveys}) and is complementary to contemporaneous larger, deeper surveys, notably 2dF \citep{colless01}, SDSS \citep{aihara11} and specially 6dFGS \citep{jones04,jones05,jones09} which also used the 2MASS XSC as its input catalog and provided a large number of redshifts for our survey. These larger surveys have not attempted to be complete over the whole sky, since many cosmological measurements do not require this level of completeness and trade-offs must be made between depth and sky coverage given available telescope time and resources. 

\subsection{Sample selection}

The initial selection of sources was based on the 2MASS Extended Source Catalog (XSC). The 2MASS photometric pipeline performed a variety of magnitude measurements for each extended source in each band. We selected as our primary set of magnitudes the isophotal magnitudes measured in an elliptical aperture defined at the K$_s=$20 mag$/\sq \arcsec$ isophote. We also include in our data tables the ``total extrapolated magnitudes'' derived by the pipeline, but do not use them for our sample selection. In the case of galaxies with angular sizes much greater than the width of a single 2MASS scan, we used 

\noindent{the photometry presented in the 2MASS Large Galaxy Atlas (LGA) by \citet{jarrett03}. We applied a modest extinction correction to the 2MASS XSC or LGA magnitudes using the maps of \citet{schlegel98}.}

We selected 45,086 sources which met the following criteria:

\begin{itemize}  
\item K$_s \leq 11.75$~mag and detected at $H$
\item $E(B-V) \leq 1$~mag
\item $|b| \geq 5^\circ$ for $30^\circ \leq l \leq 330^\circ$; $|b| \geq 8^\circ$ otherwise.
\end{itemize}

We rejected 324 sources of galactic origin (multiple stars, planetary nebulae, \ion{H}{2} regions) or pieces of galaxies detected as separate sources by the 2MASS pipeline. These rejected objects are listed in Table~\ref{tb:rej}. Additionally, we flagged 314 {\it bona fide} galaxies with compromised photometry for reprocessing at a future date. Some of these galaxies have bright stars very close to their nuclei which were not detected by the pipeline. Others are in regions of high stellar density and their center positions and/or isohpotal radii have been incorrectly measured by the pipeline. Lastly, some are close pairs or multiples but the pipeline only identified a single object.

Tom Jarrett used the original 2MASS LGA pipeline to reprocess 72 of the flagged galaxies by the date this paper was submitted for publication. These galaxies are listed in Table~\ref{tb:rep}. The remaining 242 flagged galaxies are separated in two categories. Table~\ref{tb:flg} lists 87 objects for which the photometric parameters are expected to exhibit little change after reprocessing, but would still benefit from such a procedure. These galaxies have not been removed from the catalog. Table~\ref{tb:flr} contains 165 galaxies with seriously compromised photometry, which have been removed from the catalog.

In summary, the final input catalog contains 44,599 entries which are plotted using black symbols in Figure~\ref{fig:zoa}. Galaxies outside the survey area are plotted in blue and outline the ``zone of avoidance'' described previously. In this work, we present redshifts for 43,533 of the selected galaxies, or 97.6\% of the sample.

\begin{deluxetable*}{lllcrcrr}
\tabletypesize{\scriptsize}
\tablewidth{0pc} 
\tablecaption{Telescopes and instruments used in the survey\label{tb:obs}}
\tablehead{\multicolumn{2}{l}{Observatory/Telescope} &\multicolumn{1}{l}{Camera} & \colhead{Grating} & \multicolumn{1}{c}{Coverage} & \multicolumn{1}{c}{Res.} & \multicolumn{2}{c}{N gal with $K_s$} \\ & & & \colhead{(l/mm)} & \multicolumn{1}{c}{(\AA)} & \multicolumn{1}{c}{(\AA)} & \colhead{$<11.75$}& \colhead{$>11.75$}}
\startdata 
Fred L. Whipple & 1.5-m & FAST     & 300 & 3500-7400 & 5 & 7,590 & 2,596 \\
Cerro Tololo    & 1.5-m & RCSpec   & 300 & 3700-7200 & 7 & 3,245 &   238 \\
McDonald        & 2.1-m & es2      & 600 & 3700-6400 & 4 &   114 &    50 \\
Cerro Tololo    & 4-m   & RCSpec   & 527 & 3700-7400 & 3 &    48 &       \\
Hobby-Eberly    & 9.2-m & LRS      & 300 & 4300-10800& 9 &     3 &       
\enddata
\end{deluxetable*}

\subsection{Observations, data reduction and analysis}

\begin{figure*}
\includegraphics[angle=270,width=\textwidth]{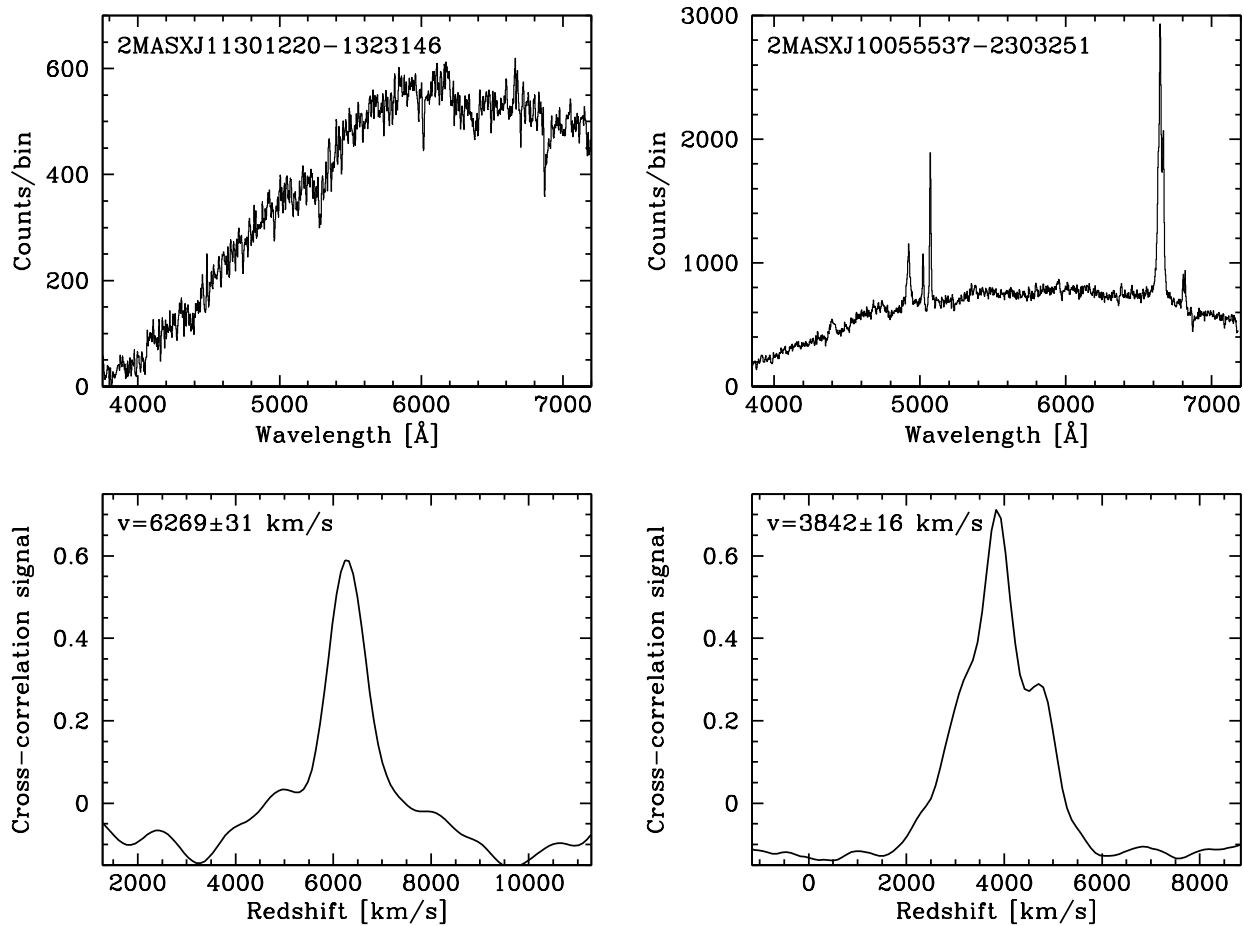}
\vspace{-3pt}
\caption{Top panels: Typical spectra obtained in this project. Left: absorption-line spectrum obtained at the FLWO 1.5-m telescope. Right: emission-line spectrum obtained at the CTIO 1.5-m telescope. Bottom panels: results of the cross-correlation technique used to measure the redshifts.\label{fig:spec}}
\end{figure*}

We obtained spectra for 11,000 galaxies that met the selection criteria listed above, plus an additional 2,898 galaxies beyond the catalog limits. Observations were carried out between September 1997 and January 2011 using a variety of facilities, listed in Table~\ref{tb:obs}. The majority of the spectra obtained for this survey were acquired at the Fred L.~Whipple Observatory 1.5-m telescope, which mostly targeted galaxies in the northern hemisphere. In the southern hemisphere, we relied heavily on observations by the 6dFGS project \citep{jones04,jones05,jones09} but also carried out our own observations using the Cerro Tololo Interamerican Observatory 1.5-m telescope. We initially targeted K$_s<11.25$~mag galaxies to obtain a complete all-sky sample \citep{huchra05} while 6dFGS observations were still ongoing. Later, we targeted galaxies below the Galactic latitude limit of 6dFGS and filled gaps in their coverage.

\vspace{4pt}

At FLWO, most observations were carried out by P.~Berlind and M.~Calkins, with additional observations by J.~Huchra, L.~Macri, A.~Crook and E.~Falco. Additional spectra were obtained in queue mode by other CfA-affiliated observers. At CTIO, observations were carried out by J.~Huchra, L.~Macri and the SMARTS consortium queue operators. At McDonald, observations were carried out by J.~Mader, T.~George and resident astronomers. Exposure times ranged from 120s to 2,400s with an average value of 550s. Some galaxies were observed on multiple nights (sometimes with increased exposure times relative to the first exposure) to improve the quality of the redshift measurement. The total ``open shutter'' time for the observations was approximately 2,100 hours. Bias and flat frames (dome or internal quartz lamp) were obtained daily. Comparison spectra were obtained before or after each science exposure using a variety of He, Ne and Ar lamps. Stellar and galaxy radial velocity standards were observed nightly.

The spectra were reduced and analyzed in a uniform manner using IRAF\footnote{IRAF is distributed by the National Optical Astronomy Observatory, which is operated by the Association of Universities for Research in Astronomy (AURA) under cooperative agreement with the National Science Foundation.}. Images were debiased and flat-fielded using routines in the {\tt CCDRED} package and one-dimensional spectra were extracted using routines in the {\tt APEXTRACT} package. Dispersion functions were derived from the comparison lamp spectra and applied to the observations using routines in the {\tt ONEDSPEC} package. The spectra obtained at FLWO were processed by S.~Tokarz and N.~Martimbeau using the automated pipeline described in \citet{tokarz97}. Two typical spectra are shown in the top panels of Figure~\ref{fig:spec}.
 
\vspace{2pt}

\begin{figure}[b]
\begin{center}
\includegraphics[angle=270,width=0.45\textwidth]{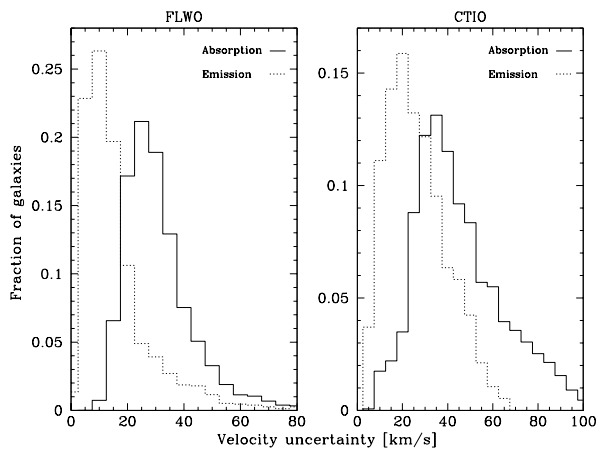}
\vspace{-9pt}
\end{center}
\caption{Distribution of velocity uncertainties for the galaxies observed at FLWO (left) and CTIO (right). The samples are further divided according to the absence or presence of emission lines.\label{fig:zerr}}
\end{figure}

Radial velocities were measured by the usual technique of cross-correlating spectra against templates \citep{tonry79} using the {\tt XCSAO} task in the {\tt RVSAO} package \citep{kurtz98}. We used a variety of templates developed at the Harvard-Smithsonian Center for Astrophysics. The bottom panels of Figure~\ref{fig:spec} show the results of the cross-correlation technique for the two representative spectra. Figure~\ref{fig:zerr} shows histograms of internal velocity uncertainties for the galaxies observed at FLWO and CTIO. The median uncertainty values for spectra that only contain absorption lines are 29 and 41 km/s for FLWO and CTIO, respectively, while the corresponding values for emission-line spectra are 12 and 24 km/s.

\vspace{2pt}

The reduced spectra are available for further analysis at the Smithsonian Astrophysical Observatory Telescope Data Center\footnote{\url{http://tdc-www.cfa.harvard.edu/2mrs/}} (hereafter, ``2MRS web site''). For example, a list of galaxies with emission-line features is available for those interested in searching for nearby AGN.

\subsection{Matching with previous redshift catalogs}

We retrieved the SDSS-DR8 spectroscopic catalog\footnote{\url{http://data.sdss3.org/sas/dr8/common/sdss-spectro/redux/ galSpecInfo-dr8.fits}} and searched for counterparts to 2MASS sources using a tolerance radius of $2\farcs5$. We found 7,069 matches to galaxies without 2MRS redshifts (including 390 galaxies with multiple SDSS observations for which we calculated a weighted mean redshift). These are identified with the catalog code ``S''.

\vspace{3pt}

We retrieved the 6dFGS-DR3 spectroscopic catalog\footnote{\url{http://www-wfau.roe.ac.uk/6dFGS/6dFGSzDR3.txt.gz}} and searched for counterparts to 2MASS sources using a tolerance radius of $10\arcsec$. We only selected redshifts measured with the 6dF instrument (code=126 in column 17 of their catalog), with velocity quality 3 or 4 (equivalent to velocity uncertainties of 55 and 45 km/s, respectively). We obtained 11,763 matches to galaxies without 2MRS redshifts. These are identified with the catalog code ``6''.

\vspace{3pt}

We performed a literature search for galaxies without 2MRS, 6dF or SDSS redshifts using the NASA Extragalactic Database (NED). First, we carried out a ``Search by Name'' query using the 2MASS IDs of the galaxies as input. This returned 12,694 redshifts that were incorporated into our catalog. We refer to these redshifts as the ``NED default'' set, and they are identified with the catalog code ``N''. Next, we performed a ``Search near Position'' query using the 2MASS coordinates of the galaxies for which no redshift information had been returned by the previous query. We used a tolerance radius of $1\farcm3$ for the search, which resulted in an additional 226 redshifts. These are galaxies where the difference in coordinates between 2MASS and previous catalogs is sufficiently large that NED has two or more entries for the same object, in most cases ``associated'' (in NED terms) with one another but no redshift information is returned when querying by 2MASS ID. In the case of an additional 32 galaxies, we did not use the default redshift returned by NED but instead adopted an alternative redshift listed in NED. These 258 ``alternative NED redshifts'' are identified with the catalog code ``M''.

\vspace{3pt}

Lastly, we matched the 2MASS XSC against John Huchra's personal compilation of redshifts (ZCAT) and found velocities for an additional 749 galaxies which had no corresponding information in NED, including 455 galaxies observed by John Huchra or collaborators prior to the start of 2MRS but never published. We also identified 77 galaxies for which the ZCAT and NED redshifts were in disagreement and we gave preference to the ZCAT values. Detailed information on these galaxies and those for which we assigned alternative NED redshifts (see preceding paragraph) is provided in Table~\ref{tb:nedz}. Galaxies with ZCAT redshifts are identified with catalog code ``O''.

\vspace{3pt}

Our catalog gives preference to 2MRS redshifts over any previously-published SDSS or 6dF value, to SDSS over 6dF, to 6dF over NED, and to NED over ZCAT, except for the cases described above. We list the additional redshifts for galaxies with multiple measurements in Table~\ref{tb:altz}, to allow interested readers to assign a different set of precedences or to compute weighted mean redshifts.

\vspace{3pt}

\begin{center}
\begin{figure}[t]
\vspace{3pt}
\includegraphics[angle=270,width=0.45\textwidth]{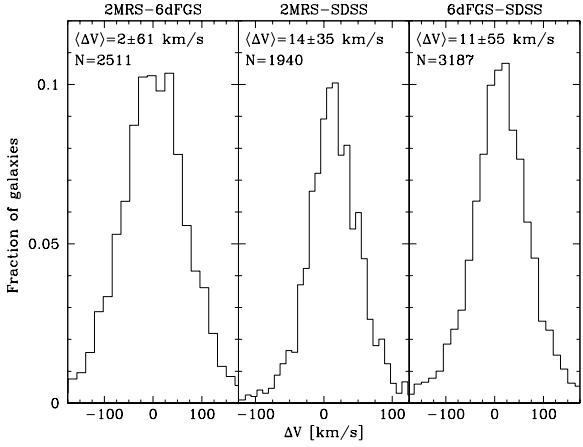}
\vspace{-3pt}
\caption{Histogram of velocity differences for galaxies observed by any two of 2MRS, 6dFGS or SDSS.\label{fig:zcomp}}
\end{figure}
\end{center}

Figure~\ref{fig:zcomp} shows a comparison of redshifts for all 2MASS galaxies observed by us and by 6dFGS or SDSS. The average redshift difference for galaxies in common between each pair of catalogs is the following: $2\pm61$~km/s for $N=2,511$~galaxies in 2MRS and 6dFGS; $14\pm35$~km/s for $N=1,940$~galaxies in 2MRS and SDSS; $11\pm55$~km/s for $N=3,187$~galaxies in 6dFGS and SDSS. The dispersions are consistent with the typical velocity uncertainties of each survey ($30-40$~km/s for 2MRS, $45-55$ km/s for 6dFGS and 5 km/s for SDSS).

\vspace*{-22pt}

\subsection{The 2MRS catalog}

The 2MRS catalog is presented in Table~\ref{tb:cat} and is also available at the 2MRS web site. It contains 29 columns that are described below, including the original 2MASS XSC column names in square brackets when applicable.

\begin{enumerate}
\item ID: 2MASS ID [designation]
\item R.A.: Right Ascension (deg, J2000.0) [sup\_ra]
\item Dec.: Declination (deg, J2000.0) [sup\_dec]
\item $l$: Galactic longitude 
\item $b$: Galactic latitude  
\item $K_s^0$: Extinction-corrected K$_s$ isophotal magnitude [k\_m\_k20fe]
\item $H^0$: same for H [h\_m\_k20fe]
\item $J^0$: same for J [j\_m\_k20fe]
\end{enumerate}

\begin{center}
\begin{figure}[t]
\includegraphics[angle=270,width=0.45\textwidth]{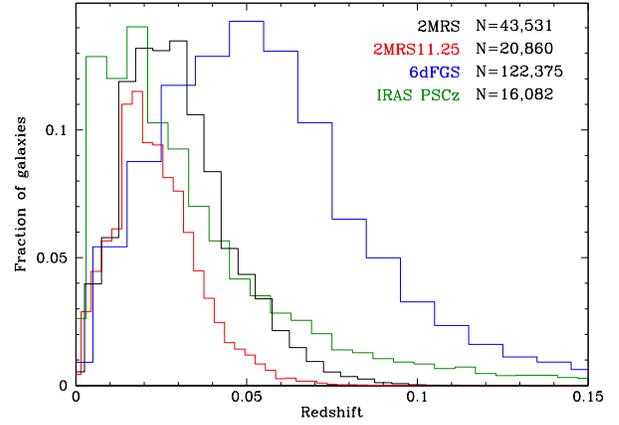}
\vspace{-9pt}
\caption{Distribution of galaxies as a function of redshift for 2MRS and for some of the redshift surveys listed in Table~\ref{tb:zsurveys}.\label{fig:zdist}}
\end{figure}
\vspace{-21pt}
\end{center}

\begin{enumerate}
\addtocounter{enumi}{8}
\item $K^0_{s,t}$: Extinction-corrected ``total'' extrapolated K$_s$ magnitude [k\_m\_ext]
\item $H^0_{t}$: same for H [h\_m\_ext]
\item $J^0_{t}$: same for J [j\_m\_ext]
\item $\sigma(K_s^0)$: Uncertainty in $K_s^0$ [k\_msig\_k20fe]
\item $\sigma(H^0)$: same for $H^0$ [h\_msig\_k20fe]
\item $\sigma(J^0)$: same for $J^0$ [j\_msig\_k20fe]
\item $\sigma(K^0_{s,t})$: same for $K^0_{t}$ [k\_msig\_ext]
\item $\sigma(H^0_{t})$: same for $H^0_{t}$ [h\_msig\_ext]
\item $\sigma(J^0_{t})$: same for $J^0_{t}$ [j\_msig\_ext]
\item $E(B-V)$: from \citet{schlegel98}
\item $r_{\rm{iso}}$: $\log_{10}$ of the K$_s=20$~mag/sq arcsec isophotal radius (in arcseconds) [r\_k20fe]
\item $r_{\rm{ext}}$: same as $r_{\rm iso}$ but for ``total magnitude'' extrapolation radius [r\_ext]
\item $b/a$: axial ratio from co-added JHK$_s$ images [sup\_ba]
\item flags: photometry confusion flags from 2MASS XSC database. "Z" in the first column indicates magnitudes from the 2MASS LGA. [cc\_flg, k\_flg\_k20fe, h\_flg\_k20fe, j\_flg\_k20fe]. 
\item type: galaxy type (see \S5 and Table~\ref{tb:types})
\item t\_src: source of galaxy type (JH=John Huchra; ZC=ZCAT; NN=not available)
\item v: redshift (km/s, barycentric)
\item $\sigma(v)$: uncertainty in redshift (km/s)
\item cat: code for redshift catalog (see notes for details).
\item v\_src: NED bibliographic code for source of redshift (see Table~\ref{tb:bib} for references)
\item Catalog ID: galaxy ID in redshift catalog
\end{enumerate}

\begin{deluxetable*}{lrrrrcccccccc}
\tabletypesize{\tiny}
\setlength{\tabcolsep}{0.02in}
\tablewidth{0pc} 
\tablecaption{2MRS Catalog (columns 1-13)\label{tb:cat}}
\tablehead{
\colhead{(1)}      & \colhead{(2)}   & \colhead{(3)}  & \colhead{(4)}   & \colhead{(5)}  & \colhead{(6)}   & \colhead{(7)}  & \colhead{(8)}   & \colhead{(9)}    & \colhead{(10)}    & \colhead{(11)}   & \colhead{(12)}          & \colhead{(13)}       \\
\colhead{2MASS ID} & \colhead{R.A.}  &\colhead{Dec.}  & \colhead{$l$}   &\colhead{$b$}   & \colhead{$K^0_s$} &\colhead{$H^0$} & \colhead{$J^0$} &\colhead{$K^0_{s,t}$} & \colhead{$H^0_t$} &\colhead{$J^0_t$} & \colhead{$\sigma(K^0_s)$} &\colhead{$\sigma(H^0)$} \\
\colhead{}         & \multicolumn{2}{c}{(deg)}        & \multicolumn{2}{c}{(deg)}        & \multicolumn{8}{c}{(mag)}}
\startdata
$00424433+4116074$ &  10.68471 &  41.26875 & 121.17430 & -21.57319 &  0.797 &  0.929 &  1.552 &  0.743 &  0.881 &  1.497 & 0.016 & 0.016 \\
$00473313-2517196$ &  11.88806 & -25.28880 &  97.36301 & -87.96452 &  3.815 &  4.132 &  4.858 &  3.765 &  4.077 &  4.798 & 0.016 & 0.015 \\
$09553318+6903549$ & 148.88826 &  69.06526 & 142.09190 &  40.90022 &  3.898 &  4.131 &  4.784 &  3.803 &  4.043 &  4.690 & 0.016 & 0.016 \\
$13252775-4301073$ & 201.36565 & -43.01871 & 309.51639 &  19.41761 &  3.948 &  4.244 &  4.931 &  3.901 &  4.203 &  4.876 & 0.015 & 0.016 \\
$13052727-4928044$ & 196.36366 & -49.46790 & 305.27151 &  13.34017 &  4.471 &  4.790 &  5.508 &  4.421 &  4.735 &  5.444 & 0.016 & 0.016 \\
$01335090+3039357$ &  23.46210 &  30.65994 & 133.61024 & -31.33081 &  4.477 &  4.697 &  5.346 &  4.087 &  4.329 &  5.003 & 0.020 & 0.018 \\
$09555243+6940469$ & 148.96846 &  69.67970 & 141.40953 &  40.56710 &  4.636 &  5.003 &  5.744 &  4.610 &  4.973 &  5.704 & 0.015 & 0.015 \\
$03464851+6805459$ &  56.70214 &  68.09611 & 138.17259 &  10.57999 &  4.682 &  4.952 &  5.494 &  4.362 &  4.682 &  5.169 & 0.020 & 0.019 \\
$13370091-2951567$ & 204.25383 & -29.86576 & 314.58353 &  31.97269 &  4.721 &  4.951 &  5.594 &  4.595 &  4.832 &  5.480 & 0.017 & 0.016 \\
$12395949-1137230$ & 189.99789 & -11.62307 & 298.46094 &  51.14923 &  4.991 &  5.228 &  5.897 &  4.944 &  5.177 &  5.841 & 0.015 & 0.015 \\
$00424182+4051546$ &  10.67427 &  40.86517 & 121.14999 & -21.97622 &  5.084 &  5.301 &  6.171 &  5.040 &  5.275 &  6.142 & 0.015 & 0.015 \\
$12505314+4107125$ & 192.72145 &  41.12015 & 123.36211 &  76.00777 &  5.163 &  5.408 &  6.068 &  5.100 &  5.344 &  6.010 & 0.015 & 0.015 \\
$12564369+2140575$ & 194.18207 &  21.68266 & 315.68127 &  84.42287 &  5.381 &  5.623 &  6.300 &  5.315 &  5.558 &  6.231 & 0.016 & 0.015 \\
$20345233+6009132$ & 308.71805 &  60.15368 &  95.71873 &  11.67289 &  5.424 &  5.921 &  6.147 &  5.248 &  5.711 &  5.971 & 0.018 & 0.017 \\
$12294679+0800014$ & 187.44499 &   8.00041 & 286.92224 &  70.19597 &  5.498 &  5.732 &  6.370 &  5.388 &  5.622 &  6.254 & 0.017 & 0.016 \\
$13295269+4711429$ & 202.46957 &  47.19526 & 104.85159 &  68.56084 &  5.589 &  5.796 &  6.486 &  5.484 &  5.632 &  6.370 & 0.017 & 0.016 \\
$12185761+4718133$ & 184.74008 &  47.30372 & 138.31985 &  68.84251 &  5.592 &  5.831 &  6.498 &  5.458 &  5.706 &  6.361 & 0.016 & 0.016 \\
$03224178-3712295$ &  50.67412 & -37.20820 & 240.16275 & -56.68984 &  5.681 &  5.947 &  6.547 &  5.580 &  5.860 &  6.427 & 0.016 & 0.015 \\
$13154932+4201454$ & 198.95554 &  42.02929 & 105.99706 &  74.28773 &  5.722 &  5.947 &  6.682 &  5.602 &  5.818 &  6.554 & 0.016 & 0.016 \\
$02424077-0000478$ &  40.66988 &  -0.01329 & 172.10397 & -51.93358 &  5.800 &  6.266 &  6.985 &  5.776 &  6.238 &  6.937 & 0.015 & 0.015 \\
$12434000+1133093$ & 190.91670 &  11.55261 & 295.87354 &  74.31767 &  5.816 &  6.064 &  6.740 &  5.730 &  5.984 &  6.647 & 0.016 & 0.016 \\
$03171859-4106290$ &  49.32750 & -41.10807 & 247.52402 & -57.04243 &  5.847 &  6.093 &  6.731 &  5.653 &  5.958 &  6.512 & 0.016 & 0.015 \\
$11054859-0002092$ & 166.45247 &  -0.03590 & 255.53194 &  52.82921 &  5.848 &  6.094 &  6.781 &  5.763 &  5.994 &  6.686 & 0.016 & 0.016 \\
$00402207+4141070$ &  10.09198 &  41.68530 & 120.71631 & -21.13871 &  5.866 &  6.099 &  6.662 &  5.557 &  5.815 &  6.374 & 0.020 & 0.019 \\
$12304942+1223279$ & 187.70593 &  12.39110 & 283.77777 &  74.49104 &  5.896 &  6.144 &  6.806 &  5.804 &  6.060 &  6.699 & 0.016 & 0.016 \\
$10051397-0743068$ & 151.30824 &  -7.71858 & 247.78252 &  36.78094 &  5.921 &  6.148 &  6.810 &  5.867 &  6.092 &  6.748 & 0.016 & 0.015 \\
$11201502+1259286$ & 170.06261 &  12.99129 & 241.96164 &  64.41846 &  5.939 &  6.195 &  6.881 &  5.869 &  6.126 &  6.806 & 0.016 & 0.016 \\
$14031258+5420555$ & 210.80243 &  54.34875 & 102.03697 &  59.77156 &  5.940 &  6.204 &  6.894 &  5.509 &  5.800 &  6.510 & 0.021 & 0.019 \\
$02223290+4220539$ &  35.63711 &  42.34832 & 140.38223 & -17.41514 &  5.971 &  6.396 &  7.279 &  5.915 &  6.324 &  7.199 & 0.016 & 0.016 \\
$04074690+6948447$ &  61.94545 &  69.81243 & 138.46324 &  13.11032 &  5.995 &  6.190 &  6.902 &  5.890 &  6.080 &  6.793 & 0.017 & 0.017 \\
$22370410+3424573$ & 339.26709 &  34.41592 &  93.72210 & -20.72380 &  6.081 &  6.332 &  7.077 &  5.996 &  6.240 &  6.978 & 0.016 & 0.015 \\
$09321011+2130029$ & 143.04213 &  21.50083 & 208.71164 &  44.54013 &  6.089 &  6.376 &  7.003 &  6.025 &  6.314 &  6.924 & 0.016 & 0.016 \\
$11185595+1305319$ & 169.73315 &  13.09221 & 241.33185 &  64.22169 &  6.104 &  6.342 &  7.018 &  6.057 &  6.294 &  6.967 & 0.016 & 0.016 \\
$12362080+2559146$ & 189.08669 &  25.98739 & 230.76158 &  86.43781 &  6.125 &  6.408 &  7.225 &  6.055 &  6.305 &  7.148 & 0.015 & 0.015 \\
$09220265+5058353$ & 140.51106 &  50.97648 & 166.94218 &  44.15141 &  6.153 &  6.381 &  7.084 &  6.057 &  6.286 &  6.993 & 0.016 & 0.016 \\
$19094609-6351271$ & 287.44208 & -63.85754 & 332.22400 & -26.14640 &  6.182 &  6.434 &  6.971 &  5.920 &  6.179 &  6.689 & 0.020 & 0.018 \\
$11201701+1335221$ & 170.07091 &  13.58949 & 240.85176 &  64.78079 &  6.211 &  6.528 &  7.351 &  6.064 &  6.361 &  7.150 & 0.016 & 0.016 \\
$12252405+1811278$ & 186.35022 &  18.19108 & 267.71216 &  79.23713 &  6.249 &  6.496 &  7.146 &  6.134 &  6.399 &  7.033 & 0.016 & 0.016 \\
$09423326-0341568$ & 145.63867 &  -3.69914 & 239.51228 &  35.01418 &  6.250 &  6.846 &  7.210 &  6.236 &  6.829 &  7.199 & 0.004 & 0.004 \\
$12261181+1256454$ & 186.54922 &  12.94597 & 279.08353 &  74.63677 &  6.273 &  6.489 &  7.165 &  6.093 &  6.310 &  6.981 & 0.017 & 0.016 \\
$02402401+3903477$ &  40.10005 &  39.06325 & 145.02322 & -19.08913 &  6.311 &  6.546 &  7.208 &  6.217 &  6.452 &  7.105 & 0.016 & 0.016 \\
$12315921+1425134$ & 187.99673 &  14.42041 & 282.32941 &  76.50777 &  6.321 &  6.604 &  7.257 &  6.254 &  6.538 &  7.178 & 0.016 & 0.016 \\
$09121949-2410213$ & 138.08124 & -24.17260 & 251.97076 &  16.35375 &  6.322 &  6.538 &  7.215 &  6.248 &  6.469 &  7.137 & 0.016 & 0.016 \\
$12250377+1253130$ & 186.26575 &  12.88696 & 278.20502 &  74.47830 &  6.333 &  6.563 &  7.224 &  6.208 &  6.445 &  7.088 & 0.016 & 0.016 \\
$04161046-5546485$ &  64.04361 & -55.78014 & 265.63147 & -43.69118 &  6.345 &  6.569 &  7.244 &  6.276 &  6.490 &  7.168 & 0.016 & 0.015 \\
$10474959+1234538$ & 161.95667 &  12.58163 & 233.49026 &  57.63276 &  6.353 &  6.573 &  7.247 &  6.261 &  6.475 &  7.147 & 0.016 & 0.015 \\
$08524134+3325184$ & 133.17227 &  33.42180 & 190.45485 &  38.76088 &  6.375 &  6.645 &  7.350 &  6.316 &  6.579 &  7.282 & 0.016 & 0.015 \\
$12502661+2530027$ & 192.61089 &  25.50076 & 295.07944 &  88.35722 &  6.380 &  6.550 &  7.342 &  6.168 &  6.301 &  7.150 & 0.017 & 0.016 \\
$13295958+4715580$ & 202.49829 &  47.26613 & 104.88158 &  68.48817 &  6.389 &  6.620 &  7.330 &  6.238 &  6.440 &  7.179 & 0.017 & 0.016 \\
$10464574+1149117$ & 161.69060 &  11.81994 & 234.43549 &  57.01038 &  6.394 &  6.633 &  7.324 &  6.311 &  6.553 &  7.214 & 0.016 & 0.016 \\
$02461905-3016296$ &  41.57941 & -30.27491 & 226.91473 & -64.68055 &  6.430 &  6.701 &  7.373 &  6.243 &  6.491 &  7.157 & 0.018 & 0.016 \\
$03382908-3527026$ &  54.62118 & -35.45074 & 236.71642 & -53.63563 &  6.436 &  6.690 &  7.353 &  6.302 &  6.552 &  7.195 & 0.017 & 0.016 \\
$07365139+6536091$ & 114.21415 &  65.60255 & 150.56915 &  29.18593 &  6.440 &  6.652 &  7.214 &  6.177 &  6.396 &  6.946 & 0.019 & 0.017 \\
$12483590-0548030$ & 192.14961 &  -5.80085 & 301.63284 &  57.06368 &  6.492 &  6.712 &  7.356 &  6.357 &  6.577 &  7.209 & 0.017 & 0.016 \\
$13105631+3703321$ & 197.73463 &  37.05894 & 101.61243 &  79.24929 &  6.496 &  6.782 &  7.505 &  6.439 &  6.724 &  7.444 & 0.015 & 0.015 \\
$12490218-0839514$ & 192.25911 &  -8.66435 & 301.91714 &  54.20262 &  6.529 &  6.777 &  7.424 &  6.489 &  6.726 &  7.367 & 0.013 & 0.008 \\
$12340302+0741569$ & 188.51262 &   7.69914 & 290.15945 &  70.13844 &  6.544 &  6.774 &  7.527 &  6.465 &  6.688 &  7.435 & 0.016 & 0.016 \\
$00145360-3911478$ &   3.72334 & -39.19663 & 332.88217 & -75.73088 &  6.557 &  6.809 &  7.328 &  6.244 &  6.538 &  6.964 & 0.021 & 0.018 \\
$12155444+1308578$ & 183.97685 &  13.14940 & 270.45587 &  73.73785 &  6.576 &  6.828 &  7.560 &  6.513 &  6.764 &  7.497 & 0.016 & 0.016 \\
$03333645-3608263$ &  53.40191 & -36.14066 & 237.95654 & -54.59778 &  6.592 &  6.922 &  7.582 &  6.366 &  6.725 &  7.345 & 0.018 & 0.017 \\
$09453879-3111279$ & 146.41164 & -31.19109 & 262.57877 &  16.76339 &  6.604 &  6.838 &  7.465 &  6.374 &  6.608 &  7.238 & 0.020 & 0.018 \\
$12374359+1149051$ & 189.43166 &  11.81809 & 290.39795 &  74.35513 &  6.608 &  6.834 &  7.469 &  6.472 &  6.683 &  7.336 & 0.018 & 0.017 \\
$11510178-2848223$ & 177.75743 & -28.80621 & 287.27591 &  32.22239 &  6.613 &  6.841 &  7.503 &  6.471 &  6.708 &  7.349 & 0.017 & 0.016 \\
$12420800+3232294$ & 190.53337 &  32.54151 & 142.80600 &  84.22353 &  6.615 &  6.967 &  7.701 &  6.459 &  6.816 &  7.510 & 0.017 & 0.017 \\
$12424986+0241160$ & 190.70778 &   2.68778 & 297.74844 &  65.47279 &  6.618 &  6.821 &  7.498 &  6.412 &  6.578 &  7.284 & 0.019 & 0.017 \\
$03385213-2620162$ &  54.71721 & -26.33784 & 221.53658 & -52.78390 &  6.687 &  6.950 &  7.591 &  6.490 &  6.769 &  7.386 & 0.017 & 0.016 \\
$05074234-3730469$ &  76.92643 & -37.51305 & 241.21202 & -35.90158 &  6.721 &  7.038 &  7.704 &  6.645 &  6.965 &  7.614 & 0.016 & 0.016 \\
$12364981+1309463$ & 189.20757 &  13.16287 & 288.46741 &  75.62270 &  6.746 &  6.957 &  7.645 &  6.565 &  6.761 &  7.462 & 0.017 & 0.016 \\
$11181630-3248453$ & 169.56792 & -32.81260 & 281.21283 &  26.09979 &  6.765 &  6.970 &  7.621 &  6.570 &  6.754 &  7.387 & 0.018 & 0.016 \\
$08372462-5507254$ & 129.35260 & -55.12374 & 271.77490 &  -8.45706 &  6.781 &  7.005 &  7.658 &  6.610 &  6.833 &  7.458 & 0.018 & 0.017 \\
$11131710-2645179$ & 168.32129 & -26.75499 & 277.24646 &  31.17530 &  6.789 &  7.013 &  7.672 &  6.680 &  6.900 &  7.554 & 0.017 & 0.016 \\
$12242822+0719030$ & 186.11761 &   7.31752 & 283.80698 &  69.18179 &  6.792 &  7.015 &  7.689 &  6.632 &  6.852 &  7.523 & 0.018 & 0.016 
\enddata
\tablecomments{This table is presented in its entirety in the online version of the paper.}
\end{deluxetable*}

\addtocounter{table}{-1}
\begin{deluxetable*}{lcccclllllrrrr}
\tabletypesize{\tiny}
\tablewidth{0pc} 
\tablecaption{2MRS Catalog (columns 14-26)}
\tablehead{
\colhead{(1)}      & \colhead{(14)}          & \colhead{(15)}            & \colhead{(16)}            & \colhead{(17)}           & \colhead{(18)}    & \colhead{(19)}        & \colhead{(20)}         & \colhead{(21)} & \colhead{(22)}  & \colhead{(23)} & \colhead{(24)}   & \colhead{(25)}   & \colhead{(26)}        \\
\colhead{2MASS ID} & \colhead{$\sigma(J^0)$} & \colhead{$\sigma(K^0_t)$} & \colhead{$\sigma(H^0_t)$} & \colhead{$\sigma(J^0_t)$} & \colhead{$E_{BV}$} &\colhead{$r_{\rm{iso}}$} & \colhead{$r_{\rm{ext}}$} &\colhead{$b/a$} & \colhead{flags} & \colhead{Type} & \colhead{t\_src} & \colhead{$v$}    & \colhead{$\sigma(v)$} \\
\colhead{}         & \multicolumn{4}{c}{(mag)}                                                                                  & \colhead{(mag)}   &\multicolumn{2}{c}{($\log_{10} \arcsec$)}        &\colhead{}      & \colhead{}      &\colhead{}      & \colhead{}       & \multicolumn{2}{c}{(km/s)}}      
\startdata
$00424433+4116074$ & 0.015 & 0.017 & 0.017 & 0.016 & 0.683 & 3.208 & 3.491 & 0.473 & Z111 &  3A2s & ZC &  -300 &   4 \\
$00473313-2517196$ & 0.015 & 0.017 & 0.016 & 0.016 & 0.019 & 2.799 & 2.965 & 0.264 & Z111 &  5X\_s & ZC &   243 &   2 \\
$09553318+6903549$ & 0.015 & 0.018 & 0.018 & 0.016 & 0.080 & 2.688 & 2.878 & 0.517 & Z111 &  2A2s & ZC &   -34 &   4 \\
$13252775-4301073$ & 0.015 & 0.016 & 0.017 & 0.016 & 0.115 & 2.445 & 2.613 & 0.957 & Z111 & -2\_\_P & ZC &   547 &   5 \\
$13052727-4928044$ & 0.015 & 0.017 & 0.017 & 0.016 & 0.176 & 2.627 & 2.772 & 0.308 & Z111 &  6B\_s & ZC &   563 &   3 \\
$01335090+3039357$ & 0.017 & 0.044 & 0.038 & 0.029 & 0.041 & 2.699 & 3.032 & 0.792 & Z111 &  5A4s & ZC &  -179 &   3 \\
$09555243+6940469$ & 0.015 & 0.015 & 0.015 & 0.015 & 0.156 & 2.357 & 2.542 & 0.396 & Z111 &  0    & ZC &   203 &   4 \\
$03464851+6805459$ & 0.018 & 0.043 & 0.040 & 0.033 & 0.558 & 2.571 & 2.876 & 0.858 & Z111 &  6X2T & ZC &    31 &   3 \\
$13370091-2951567$ & 0.016 & 0.025 & 0.019 & 0.018 & 0.067 & 2.495 & 2.709 & 0.825 & Z111 &  5X2s & ZC &   513 &   2 \\
$12395949-1137230$ & 0.015 & 0.017 & 0.015 & 0.015 & 0.051 & 2.305 & 2.473 & 0.682 & Z111 &  1A\_P & ZC &  1024 &   5 \\
$00424182+4051546$ & 0.015 & 0.017 & 0.016 & 0.016 & 0.155 & 2.168 & 2.360 & 0.913 & Z111 & -6    & ZC &  -200 &   6 \\
$12505314+4107125$ & 0.015 & 0.016 & 0.016 & 0.015 & 0.018 & 2.236 & 2.414 & 0.847 & Z111 &  2A3R & ZC &   308 &   1 \\
$12564369+2140575$ & 0.015 & 0.017 & 0.016 & 0.016 & 0.041 & 2.332 & 2.490 & 0.583 & Z111 &  2A\_T & ZC &   408 &   4 \\
$20345233+6009132$ & 0.017 & 0.034 & 0.029 & 0.025 & 0.342 & 2.402 & 2.680 & 0.770 & Z111 &  6X1T & ZC &    40 &   2 \\
$12294679+0800014$ & 0.016 & 0.025 & 0.021 & 0.019 & 0.022 & 2.253 & 2.496 & 0.913 & Z333 & -5    & ZC &   997 &   7 \\
$13295269+4711429$ & 0.016 & 0.025 & 0.020 & 0.019 & 0.035 & 2.296 & 2.549 & 0.902 & Z111 &  4A1P & ZC &   463 &   3 \\
$12185761+4718133$ & 0.016 & 0.022 & 0.021 & 0.018 & 0.016 & 2.421 & 2.662 & 0.495 & Z333 &  4X\_s & ZC &   448 &   3 \\
$03224178-3712295$ & 0.015 & 0.019 & 0.018 & 0.017 & 0.021 & 2.220 & 2.470 & 0.792 & Z133 & -2X\_P & ZC &  1760 &  10 \\
$13154932+4201454$ & 0.016 & 0.021 & 0.020 & 0.018 & 0.018 & 2.310 & 2.541 & 0.660 & Z000 &  4A3T & ZC &   484 &   1 \\
$02424077-0000478$ & 0.015 & 0.016 & 0.015 & 0.015 & 0.033 & 1.978 & 2.162 & 0.880 & Z111 &  3A\_T & ZC &  1137 &   3 \\
$12434000+1133093$ & 0.015 & 0.021 & 0.017 & 0.017 & 0.026 & 2.166 & 2.383 & 0.891 & Z111 & -5    & ZC &  1117 &   6 \\
$03171859-4106290$ & 0.015 & 0.024 & 0.021 & 0.019 & 0.013 & 2.157 & 2.513 & 0.891 & Z111 &  0B\_s & ZC &   788 &  45 \\
$11054859-0002092$ & 0.015 & 0.021 & 0.018 & 0.017 & 0.057 & 2.216 & 2.432 & 0.693 & Z111 &  4X3T & ZC &   801 &   3 \\
$00402207+4141070$ & 0.017 & 0.045 & 0.040 & 0.029 & 0.085 & 2.450 & 2.760 & 0.594 & Z111 & -5    & ZC &  -241 &   3 \\
$12304942+1223279$ & 0.015 & 0.019 & 0.018 & 0.017 & 0.023 & 2.134 & 2.368 & 0.990 & Z111 & -4\_\_P & ZC &  1307 &   7 \\
$10051397-0743068$ & 0.015 & 0.017 & 0.016 & 0.016 & 0.046 & 2.216 & 2.397 & 0.451 & Z111 & -3    & ZC &   663 &   4 \\
$11201502+1259286$ & 0.015 & 0.018 & 0.017 & 0.016 & 0.033 & 2.267 & 2.447 & 0.539 & Z111 &  3X3s & ZC &   727 &   3 \\
$14031258+5420555$ & 0.017 & 0.050 & 0.042 & 0.032 & 0.009 & 2.373 & 2.721 & 0.979 & Z111 &  6X1T & ZC &   246 &   4 \\
$02223290+4220539$ & 0.016 & 0.017 & 0.017 & 0.017 & 0.065 & 2.353 & 2.523 & 0.264 & Z111 &  5A\_s & ZC &   528 &   4 \\
$04074690+6948447$ & 0.016 & 0.019 & 0.019 & 0.017 & 0.421 & 2.200 & 2.355 & 0.803 & Z111 &  2A\_s & ZC &   895 &   1 \\
$22370410+3424573$ & 0.015 & 0.018 & 0.017 & 0.016 & 0.091 & 2.193 & 2.425 & 0.539 & Z111 &  4A2s & ZC &   816 &   1 \\
$09321011+2130029$ & 0.015 & 0.018 & 0.017 & 0.016 & 0.031 & 2.212 & 2.393 & 0.616 & Z133 &  4X2T & ZC &   556 &   1 \\
$11185595+1305319$ & 0.015 & 0.017 & 0.017 & 0.016 & 0.025 & 2.326 & 2.457 & 0.308 & Z111 &  1X3T & ZC &   807 &   3 \\
$12362080+2559146$ & 0.015 & 0.017 & 0.017 & 0.016 & 0.015 & 2.281 & 2.513 & 0.330 & Z111 &  3A1s & ZC &  1230 &   5 \\
$09220265+5058353$ & 0.015 & 0.019 & 0.017 & 0.016 & 0.015 & 2.258 & 2.459 & 0.495 & Z111 &  3A1R & ZC &   638 &   3 \\
$19094609-6351271$ & 0.017 & 0.047 & 0.039 & 0.028 & 0.043 & 2.337 & 2.663 & 0.726 & Z111 &  4X3R & ZC &   841 &   2 \\
$11201701+1335221$ & 0.016 & 0.023 & 0.023 & 0.020 & 0.027 & 2.403 & 2.680 & 0.286 & Z111 &  3\_\_P & ZC &   825 &   2 \\
$12252405+1811278$ & 0.015 & 0.021 & 0.020 & 0.017 & 0.030 & 2.170 & 2.399 & 0.781 & Z111 & -1A\_P & ZC &   729 &   2 \\
$09423326-0341568$ & 0.002 & 0.005 & 0.004 & 0.002 & 0.054 & 1.852 & 1.997 & 0.560 & 0000 & -5    & ZC &  1919 &  13 \\
$12261181+1256454$ & 0.016 & 0.028 & 0.026 & 0.021 & 0.029 & 2.180 & 2.486 & 0.825 & Z113 & -5    & ZC &  -244 &   5 \\
$02402401+3903477$ & 0.015 & 0.021 & 0.018 & 0.017 & 0.061 & 2.258 & 2.494 & 0.374 & Z111 & -2B\_T & ZC &   637 &   4 \\
$12315921+1425134$ & 0.015 & 0.017 & 0.017 & 0.016 & 0.038 & 2.191 & 2.363 & 0.506 & Z111 &  3A1T & ZC &  2281 &   3 \\
$09121949-2410213$ & 0.015 & 0.018 & 0.017 & 0.016 & 0.210 & 2.170 & 2.331 & 0.484 & Z111 & -2A\_s & ZC &   686 &  45 \\
$12250377+1253130$ & 0.015 & 0.023 & 0.022 & 0.018 & 0.041 & 2.061 & 2.358 & 1.000 & Z111 & -5    & ZC &  1060 &   6 \\
$04161046-5546485$ & 0.015 & 0.018 & 0.017 & 0.016 & 0.015 & 2.066 & 2.242 & 0.748 & Z111 & -2A\_R & ZC &  1201 &  16 \\
$10474959+1234538$ & 0.015 & 0.018 & 0.017 & 0.016 & 0.025 & 2.033 & 2.283 & 0.979 & Z111 & -5    & ZC &   911 &   2 \\
$08524134+3325184$ & 0.015 & 0.017 & 0.016 & 0.016 & 0.033 & 2.185 & 2.362 & 0.396 & Z111 &  3A3t & ZC &   411 &   1 \\
$12502661+2530027$ & 0.016 & 0.031 & 0.026 & 0.024 & 0.012 & 2.242 & 2.562 & 0.704 & Z111 &  2X1R & ZC &  1206 &   3 \\
$13295958+4715580$ & 0.016 & 0.030 & 0.023 & 0.021 & 0.035 & 2.094 & 2.416 & 0.990 & Z111 &  0\_\_P & ZC &   465 &  10 \\
$10464574+1149117$ & 0.015 & 0.023 & 0.019 & 0.018 & 0.025 & 2.086 & 2.336 & 0.726 & Z111 &  2X\_T & ZC &   897 &   4 \\
$02461905-3016296$ & 0.016 & 0.034 & 0.025 & 0.022 & 0.027 & 2.198 & 2.513 & 0.605 & Z111 &  3B2s & ZC &  1271 &   3 \\
$03382908-3527026$ & 0.015 & 0.027 & 0.021 & 0.019 & 0.013 & 2.055 & 2.343 & 1.000 & Z111 & -5    & ZC &  1425 &   4 \\
$07365139+6536091$ & 0.016 & 0.039 & 0.028 & 0.023 & 0.040 & 2.262 & 2.576 & 0.781 & Z111 &  6X5s & ZC &   131 &   3 \\
$12483590-0548030$ & 0.015 & 0.027 & 0.020 & 0.019 & 0.029 & 2.093 & 2.381 & 0.803 & Z111 & -5    & ZC &  1241 &   2 \\
$13105631+3703321$ & 0.015 & 0.016 & 0.016 & 0.016 & 0.014 & 2.117 & 2.296 & 0.484 & Z111 &  3A3s & ZC &   946 &   5 \\
$12490218-0839514$ & 0.007 & 0.013 & 0.008 & 0.007 & 0.034 & 2.004 & 2.139 & 0.800 & 0000 &  3X\_T & ZC &  1277 &  45 \\
$12340302+0741569$ & 0.016 & 0.020 & 0.018 & 0.017 & 0.022 & 2.179 & 2.395 & 0.385 & Z000 & -2X\_s & ZC &   448 &   8 \\
$00145360-3911478$ & 0.016 & 0.049 & 0.036 & 0.025 & 0.013 & 2.437 & 2.760 & 0.330 & Z111 &  9B\_s & ZC &   129 &   2 \\
$12155444+1308578$ & 0.015 & 0.018 & 0.017 & 0.016 & 0.032 & 2.338 & 2.502 & 0.198 & Z111 &  3X3s & ZC &   131 &   4 \\
$03333645-3608263$ & 0.017 & 0.036 & 0.031 & 0.026 & 0.020 & 2.132 & 2.441 & 0.825 & Z111 &  3B2s & ZC &  1664 &  45 \\
$09453879-3111279$ & 0.017 & 0.043 & 0.031 & 0.029 & 0.108 & 2.234 & 2.530 & 0.726 & Z111 &  5X1T & ZC &  1088 &   2 \\
$12374359+1149051$ & 0.016 & 0.031 & 0.025 & 0.021 & 0.041 & 2.079 & 2.358 & 0.946 & Z111 &  3X\_T & ZC &  1517 &   1 \\
$11510178-2848223$ & 0.016 & 0.030 & 0.022 & 0.021 & 0.081 & 2.078 & 2.363 & 0.814 & Z111 & -5    & ZC &  1807 &  45 \\
$12420800+3232294$ & 0.016 & 0.024 & 0.025 & 0.020 & 0.017 & 2.348 & 2.617 & 0.264 & Z113 &  7B5/ & JH &   606 &   3 \\
$12424986+0241160$ & 0.016 & 0.035 & 0.027 & 0.023 & 0.029 & 2.130 & 2.416 & 0.836 & Z111 & -5    & ZC &   938 &   4 \\
$03385213-2620162$ & 0.016 & 0.034 & 0.026 & 0.022 & 0.013 & 2.054 & 2.381 & 0.990 & Z111 &  2B1R & ZC &  1297 &  45 \\
$05074234-3730469$ & 0.015 & 0.021 & 0.018 & 0.017 & 0.030 & 2.118 & 2.341 & 0.484 & Z111 &  1X\_s & ZC &   968 &  45 \\
$12364981+1309463$ & 0.016 & 0.027 & 0.024 & 0.020 & 0.047 & 2.218 & 2.524 & 0.484 & Z000 &  2X\_T & ZC &  -235 &   4 \\
$11181630-3248453$ & 0.016 & 0.036 & 0.024 & 0.023 & 0.081 & 2.213 & 2.513 & 0.528 & Z111 &  7A6s & ZC &   720 &  45 \\
$08372462-5507254$ & 0.016 & 0.027 & 0.023 & 0.021 & 0.298 & 2.034 & 2.289 & 0.880 & Z111 & -2    & ZC &  1051 &  32 \\
$11131710-2645179$ & 0.016 & 0.021 & 0.018 & 0.017 & 0.064 & 2.045 & 2.261 & 0.693 & Z111 & -5    & ZC &  1367 &  45 \\
$12242822+0719030$ & 0.016 & 0.029 & 0.022 & 0.020 & 0.021 & 2.053 & 2.312 & 0.836 & Z000 & -5    & ZC &  1243 &   6 
\enddata
\tablecomments{This table is presented in its entirety in the online version of the paper.}
\end{deluxetable*}

\addtocounter{table}{-1}
\begin{deluxetable*}{lcll}
\tabletypesize{\tiny}
\tablewidth{0pc} 
\tablecaption{2MRS Catalog (columns 27-29)}
\tablehead{
\colhead{(1)}      & \colhead{(27)} & \colhead{(28)}     & \colhead{(29)}      \\
\colhead{2MASS ID} & \colhead{cat}  & \colhead{Bibcode}  & \colhead{Catalog ID}}
\startdata
$00424433+4116074$ & N & 1991RC3.9.C...0000d & MESSIER 031                 \\
$00473313-2517196$ & N & 2004AJ....128...16K & NGC 0253                    \\
$09553318+6903549$ & N & 1991RC3.9.C...0000d & MESSIER 081                 \\
$13252775-4301073$ & N & 1978PASP...90..237G & NGC 5128                    \\
$13052727-4928044$ & N & 2004AJ....128...16K & NGC 4945                    \\
$01335090+3039357$ & N & 1991RC3.9.C...0000d & MESSIER 033                 \\
$09555243+6940469$ & N & 1991RC3.9.C...0000d & MESSIER 082                 \\
$03464851+6805459$ & N & 1999PASP..111..438F & IC 0342                     \\
$13370091-2951567$ & N & 2004AJ....128...16K & MESSIER 083                 \\
$12395949-1137230$ & N & 2000MNRAS.313..469S & MESSIER 104                 \\
$00424182+4051546$ & N & 2000UZC...C......0F & MESSIER 032                 \\
$12505314+4107125$ & N & 1993A\&A...272...63M& MESSIER 094                 \\
$12564369+2140575$ & N & 1991RC3.9.C...0000d & MESSIER 064                 \\
$20345233+6009132$ & N & 2008MNRAS.388..500E & NGC 6946                    \\
$12294679+0800014$ & N & 2000MNRAS.313..469S & MESSIER 049                 \\
$13295269+4711429$ & N & 1991RC3.9.C...0000d & MESSIER 051a                \\
$12185761+4718133$ & N & 1991RC3.9.C...0000d & MESSIER 106                 \\
$03224178-3712295$ & N & 1998A\&AS..130..267L& NGC 1316                    \\
$13154932+4201454$ & N & 2008MNRAS.388..500E & MESSIER 063                 \\
$02424077-0000478$ & N & 1999ApJS..121..287H & MESSIER 077                 \\
$12434000+1133093$ & N & 2000AJ....119.1645T & MESSIER 060                 \\
$03171859-4106290$ & 6 & 20096dF...C...0000J & g0317186-410629             \\
$11054859-0002092$ & N & 2004AJ....128...16K & NGC 3521                    \\
$00402207+4141070$ & N & 1991A\&A...246..349B& MESSIER 110                 \\
$12304942+1223279$ & N & 2000MNRAS.313..469S & MESSIER 087                 \\
$10051397-0743068$ & N & 2006MNRAS.367..815N & NGC 3115                    \\
$11201502+1259286$ & N & 1991RC3.9.C...0000d & MESSIER 066                 \\
$14031258+5420555$ & S & 2011SDSS8.C...0000: & 14031253+5420561            \\
$02223290+4220539$ & N & 1991RC3.9.C...0000d & NGC 0891                    \\
$04074690+6948447$ & N & 1988ApJS...67....1T & IC 0356                     \\
$22370410+3424573$ & N & 1998AJ....115...62H & NGC 7331                    \\
$09321011+2130029$ & N & 1998AJ....115...62H & NGC 2903                    \\
$11185595+1305319$ & N & 1991RC3.9.C...0000d & MESSIER 065                 \\
$12362080+2559146$ & N & 1994PASJ...46..147S & NGC 4565                    \\
$09220265+5058353$ & N & 1991RC3.9.C...0000d & NGC 2841                    \\
$19094609-6351271$ & N & 2004AJ....128...16K & NGC 6744                    \\
$11201701+1335221$ & S & 2011SDSS8.C...0000: & 11201701+1335228            \\
$12252405+1811278$ & N & 2000MNRAS.313..469S & MESSIER 085                 \\
$09423326-0341568$ & N & 2008AJ....135.2424O & NGC 2974                    \\
$12261181+1256454$ & N & 2000MNRAS.313..469S & MESSIER 086                 \\
$02402401+3903477$ & N & 1991RC3.9.C...0000d & NGC 1023                    \\
$12315921+1425134$ & N & 1985AJ.....90.1681B & MESSIER 088                 \\
$09121949-2410213$ & 6 & 20096dF...C...0000J & g0912194-241021             \\
$12250377+1253130$ & N & 2000AJ....119.1645T & MESSIER 084                 \\
$04161046-5546485$ & C & 20112MRS.CTIO.0000H & 04161030-5546510            \\
$10474959+1234538$ & N & 2000MNRAS.313..469S & MESSIER 105                 \\
$08524134+3325184$ & N & 1998AJ....115...62H & NGC 2683                    \\
$12502661+2530027$ & N & 1991RC3.9.C...0000d & NGC 4725                    \\
$13295958+4715580$ & N & 1991RC3.9.C...0000d & MESSIER 051b                \\
$10464574+1149117$ & N & 1991RC3.9.C...0000d & MESSIER 096                 \\
$02461905-3016296$ & N & 2004AJ....128...16K & NGC 1097                    \\
$03382908-3527026$ & N & 1998A\&AS..133..325G& NGC 1399                    \\
$07365139+6536091$ & N & 1991RC3.9.C...0000d & NGC 2403                    \\
$12483590-0548030$ & N & 2000MNRAS.313..469S & NGC 4697                    \\
$13105631+3703321$ & N & 1991RC3.9.C...0000d & NGC 5005                    \\
$12490218-0839514$ & 6 & 20096dF...C...0000J & g1249022-083952             \\
$12340302+0741569$ & N & 1991RC3.9.C...0000d & NGC 4526                    \\
$00145360-3911478$ & N & 2004AJ....128...16K & NGC 0055                    \\
$12155444+1308578$ & N & 1991RC3.9.C...0000d & NGC 4216                    \\
$03333645-3608263$ & 6 & 20096dF...C...0000J & g0333364-360826             \\
$09453879-3111279$ & N & 2004AJ....128...16K & NGC 2997                    \\
$12374359+1149051$ & N & 2008AJ....136..713K & MESSIER 058                 \\
$11510178-2848223$ & 6 & 20096dF...C...0000J & g1151017-284821             \\
$12420800+3232294$ & N & 1991RC3.9.C...0000d & NGC 4631                    \\
$12424986+0241160$ & N & 2000MNRAS.313..469S & NGC 4636                    \\
$03385213-2620162$ & 6 & 20096dF...C...0000J & g0338521-262016             \\
$05074234-3730469$ & 6 & 20096dF...C...0000J & g0507423-373046             \\
$12364981+1309463$ & N & 1991RC3.9.C...0000d & MESSIER 090                 \\
$11181630-3248453$ & 6 & 20096dF...C...0000J & g1118165-324851             \\
$08372462-5507254$ & N & 1992ApJS...83...29S & NGC 2640                    \\
$11131710-2645179$ & 6 & 20096dF...C...0000J & g1113171-264518             \\
$12242822+0719030$ & N & 2000MNRAS.313..469S & NGC 4365                    
\enddata
\tablecomments{This table is presented in its entirety in the online version of the paper.\\Codes for column 27: [C]TIO, Mc[D]onald, [F]LWO, [N]ED 2MASS ID match,\\NED position [M]atch, [O]ther sources in ZCAT, [S]DSS-DR8, [6]dFGS.}
\end{deluxetable*}

\vspace{4pt}

In addition to our measurements, Table~\ref{tb:cat} contains redshifts from 578 publications which are referenced in the catalog using ADS/NED bibliographic codes (see Table~\ref{tb:bib}). We strongly encourage proper citation of the original publications when making use of any of these values.
 
\vspace{4pt}

Table~\ref{tb:zextra} lists 4,291 redshifts for 2MASS galaxies which lie beyond the limits of our main catalog. 2,884 were observed as part of this project while 1,407 had been previously targeted by Huchra and collaborators for other projects. Lastly, Table~\ref{tb:znocat} presents redshifts for 14 galaxies that are not in the 2MASS XSC but which were observed serendipitously due to their proximity to our targets.

\vspace{4pt}

Figure~\ref{fig:zdist} shows the distribution of galaxies as a function of redshift for the 2MRS main sample and selected surveys from Table~\ref{tb:zsurveys}.

\begin{figure}[t]
\begin{center}
\includegraphics[width=0.45\textwidth]{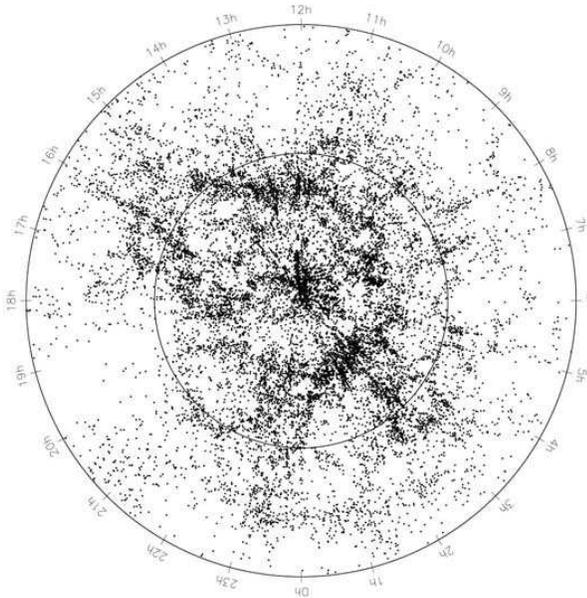}
\end{center}
\caption{Hockey Puck plot --- a full cylinder section --- of 2MRS in the north celestial cap.  The view is looking downwards from the NCP, the thickness of the ``puck'' is 8,000 km/s and its radius is 15,000 km/s. \label{fig:hp1}}
\vspace{9pt}
\end{figure}

\vspace{9pt}

\section{Cosmic Cartography}

Some initial qualitative results from this survey are shown below via two visualization techniques: Hockey Pucks and Onion Skins.

\subsection{Hockey Pucks}

An all-sky survey allows us to make plots of the nearby galaxy distribution that are more representative than simple strip surveys \citep{delapparent86}. The angular nature of strips around the sky, when projected onto a plane, are somewhat deceptive of real structure.  They are thin at the center and thick at the edge. While this partially makes up for the normal decrease in the selection efficiency as a function of redshift in a flux limited sample, it provides a representation of structure that varies quite strongly from the center to edge. With full sky coverage, it is possible to project actual cylinders of redshift space. Given the long-term association of redshift surveys with the Harvard-Smithsonian CfA we naturally call these ``Hockey Puck'' plots. Code to generate these plots is available as part of the 2MRS data release.

\vspace{6pt}
 
\begin{figure}[t]
\begin{center}
\includegraphics[width=0.45\textwidth]{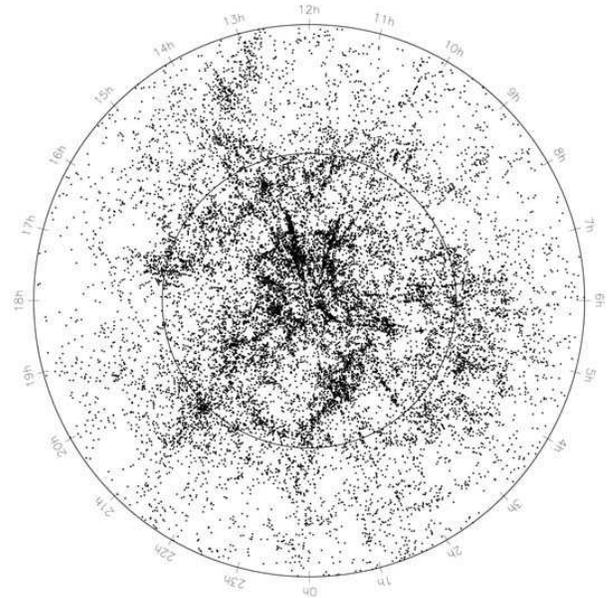}
\end{center}
\caption{Same as Fig.~\ref{fig:hp1} but for the south celestial cap.\label{fig:hp2}}
\end{figure}

Two ``Hockey puck'' diagrams shown in Figures~\ref{fig:hp1} and \ref{fig:hp2} highlight the vast improvement in coverage through the galactic plane afforded by 2MRS as compared to even CfA2, the densest survey of the nearby universe \citep{huchra95,huchra99}.  Plotted are top-down views of cylindrical volumes with a radius of 15000 km/s and thickness of 8000 km/s, yielding an aspect ratio of about 3.5 to 1. The pucks show the galaxies in the northern and southern celestial hemispheres respectively -- i.e. all galaxies above and below the celestial equator with redshifts placing them in the cylinder and with K$_s \leq$ 11.75~mag. Many of our favorite structures and several prominent voids are easily seen in these plots.

\begin{figure*}
\begin{center}
\includegraphics[height=2.5in]{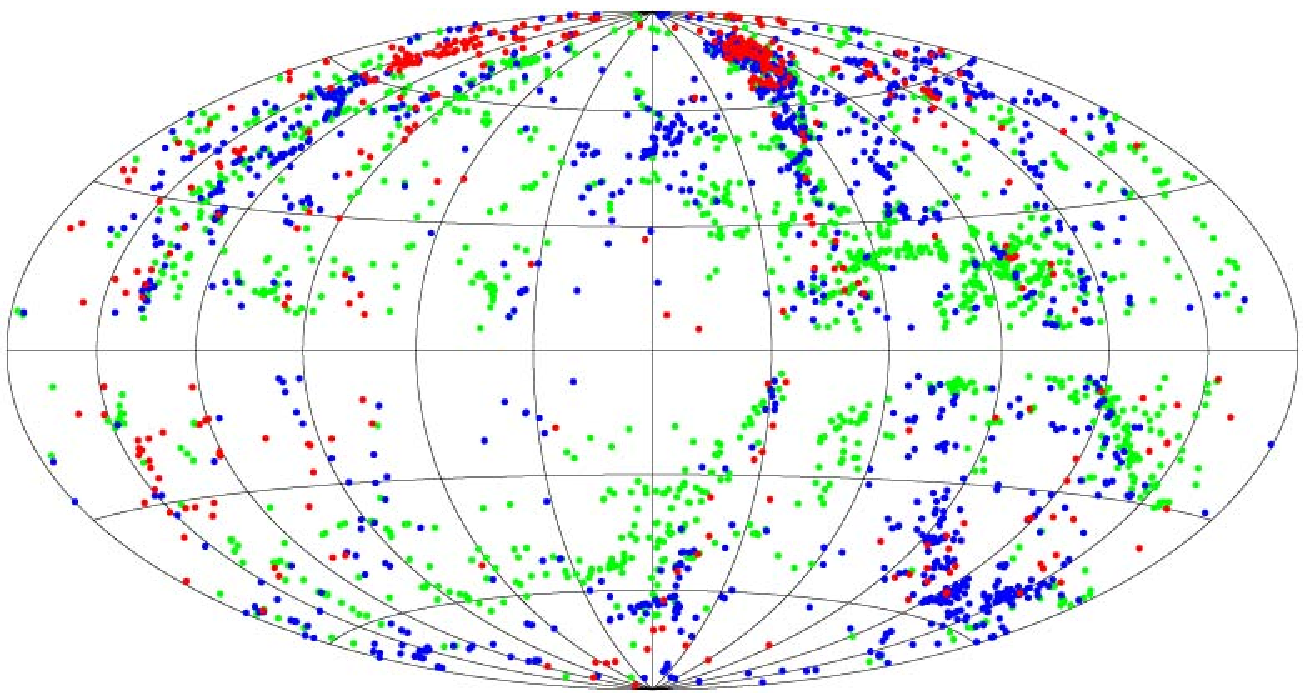}
\end{center}
\caption{2MASS galaxies inside the 3,000 km/s sphere in Galactic coordinates (centered at $l=0^\circ$ and following the convention of $l$ increasing to the left). Heliocentric velocities are color coded with red, blue \& green representing bins of increasing redshift/distance.  Red for $V_h < 1,000$~km/s, blue for $1,000 < V_h < 2,000$~km/s, and green for $2,000 < V_h < 3,000$~km/s.\label{fig:os1}} 
\end{figure*}

\begin{figure*}
\begin{center}
\includegraphics[height=2.5in]{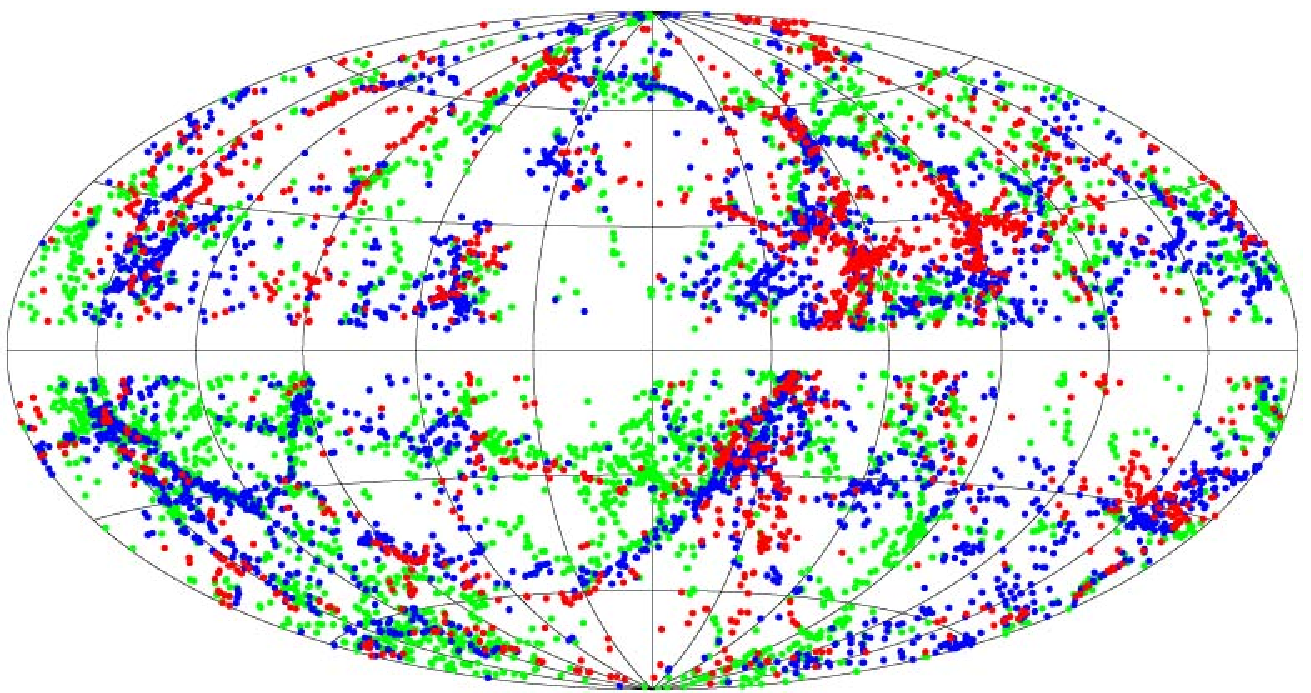}
\end{center}
\caption{Same as Fig.~\ref{fig:os1}, but for velocities between 3,000 and 6,000 km/s. \label{fig:os2}}
\end{figure*}

\begin{figure*}
\begin{center}
\includegraphics[height=2.5in]{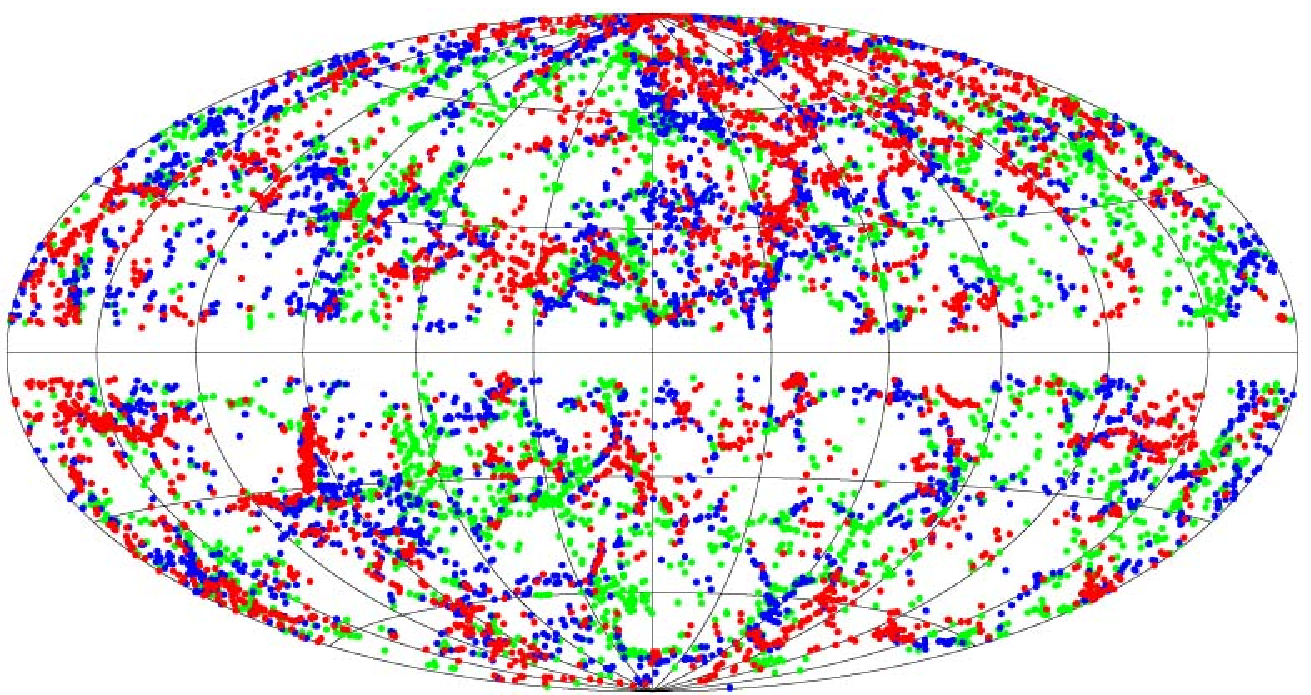}
\end{center}
\caption{Same as Fig.~\ref{fig:os1}, but for velocities between 6,000 and 9,000 km/s. \label{fig:os3}}
\end{figure*}

\begin{figure*}[t]
\begin{center}
\includegraphics[angle=90,height=1\textheight]{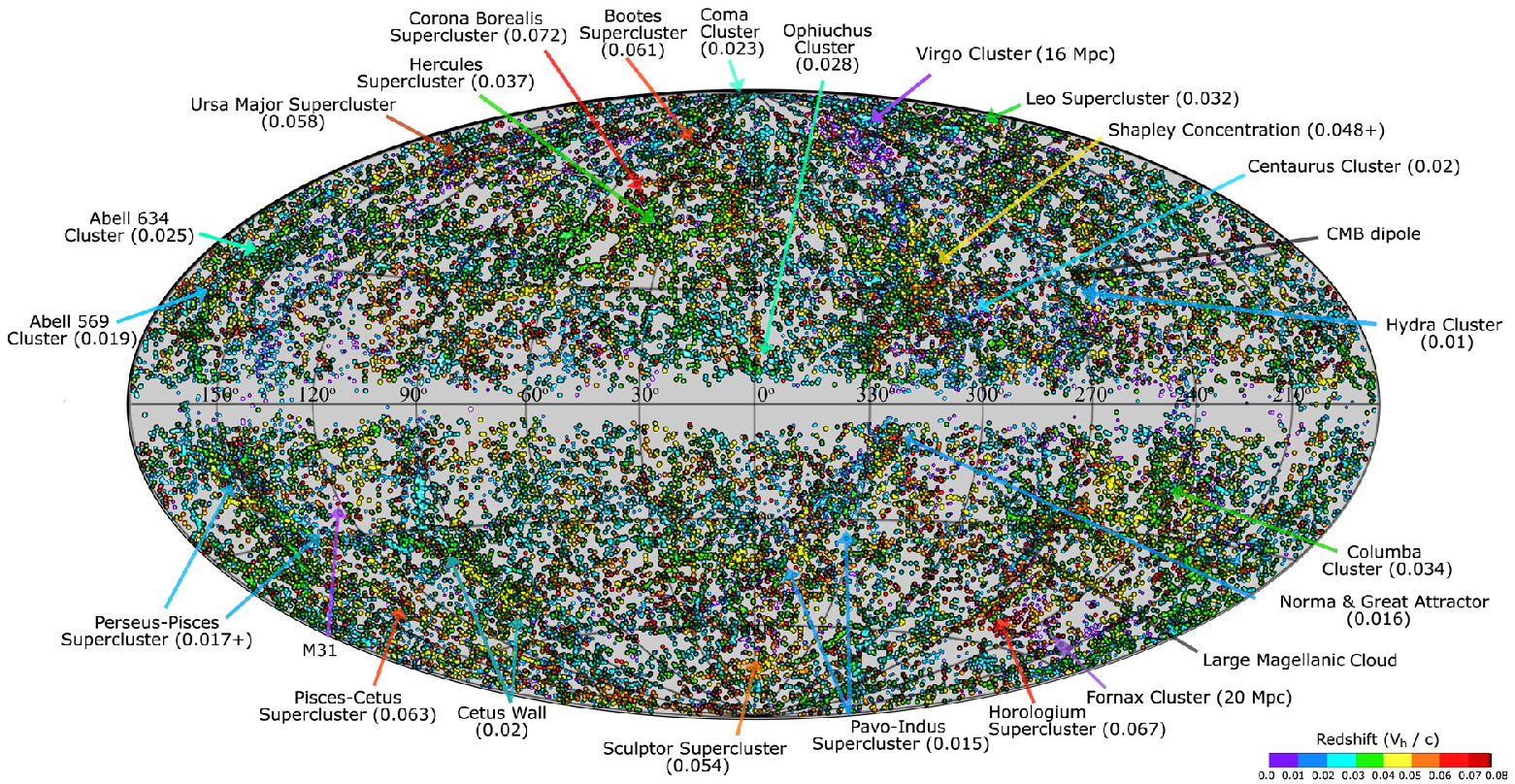}
\end{center}
\caption{Same as Fig.~\ref{fig:os1} but for all 2MRS galaxies, spanning the entire redshift range covered by the survey (from $z=0$ in purple to $z=0.08$ in red). \label{fig:os4}}
\end{figure*}

\vspace{6pt}
 
The northern puck is dominated by the Local Supercluster at the center, the Great Wall (now straight in this cylindrical projection!) at 10-14.5 hours and Pisces-Perseus at 0-5 hours.  In addition, there are several new but smaller structures such as the one at 19 hours and 4,000 km/s, probably best associated with the Cygnus Cluster \citep{huchra77}.

\vspace{6pt}

The southern celestial hemisphere is more amorphous.  There is the well-known Cetus Wall \citep{fairall98} between 0 and 4 hours, the southern part of the Local Supercluster at the center, and the Hydra-Centaurus region, but also a large and diffuse overdensity between 19 and 22 hours, a region hitherto not mapped because of its proximity to the galactic plane.  This structure appears to be both large and rich and should have a large effect on the local velocity field.

\vspace{9pt}

\subsection{Onion Skins}

Another projection that can highlight the properties of nearby structures are surface maps of the galaxy distribution as a function of redshift.  Since these are conceptually like peeling an onion, they are best called ``onion skins.''  Figures~\ref{fig:os1}-\ref{fig:os3} show three sets of these skins, moving progressively outward in redshift, while Figure~\ref{fig:os4} shows the entire 2MRS catalog with the major structures of the Local Universe labeled. These figures use Galactic coordinate projections; the corresponding equatorial coordinate projections are shown in the Appendix.

\vfill\pagebreak

\begin{center}
\begin{figure}[t]
\includegraphics[angle=270,width=0.45\textwidth]{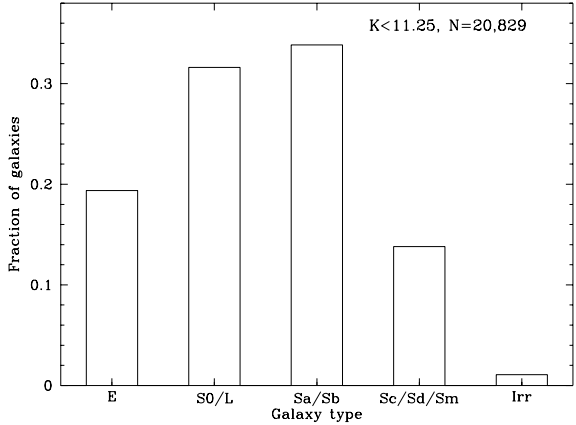}
\caption{Histogram of the distribution of galaxy types for the K$_s<11.25$~mag, $|b|>10^\circ$ sample. \label{fig:histtype}}
\end{figure}
\end{center}

Figure~\ref{fig:os1} shows the distribution on the sky of all galaxies in the survey inside 3000 km/s color coded by redshift in 1000 km/s skins.  The plane of the Local Supercluster dominates the map, but there is also a diffuse component between 2000 and 3000 km/s and 6 to 13 hours in the south. The next two figures again show some familiar structures but with a few surprises. The Great Wall, Pisces-Perseus and the Great Attractor dominate the mid ranges.  The overdensity of galaxies in the direction of A3627 is high, and the comparison of Figure~\ref{fig:os2} with \ref{fig:os3} clearly shows why we are moving with respect to the CMB towards a point around $l=270^\circ,b=30^\circ$.

\section{Galaxy Morphologies}
Morphological types are listed in Table~\ref{tb:cat} for all of the 20,860~galaxies in 2MRS11.25. We used the classifications listed in ZCAT (based on RC3, NED and other catalogs) when available, but 5,682 of these galaxies had no type information. They were visually examined and classified by one of us (J.P.H.) using blue plates from the Digitized Sky Surveys. These new morphological types are identified by code ``JH'' in column 24 of the catalog. We also list morphological types from the literature for fainter galaxies in the catalog, when available.

\vspace{3pt}

Morphological typing in 2MRS uses the modified Hubble sequence \citep{devaucouleurs63,devaucouleurs76}. Elliptical galaxies have integer types $-7$ through $-5$. S0 galaxies range from integer type $-4$ (E/S0) through 0 (S0/a), in a sequence from least to most pronounced disks. Spirals are assigned integer types 1 (Sa) through 9 (Sm), without distinction between barred, unbarred or mixed-type. Irregular and peculiar galaxies are assigned integer types 10 and above. The format for the morphological type designations is described in detail in Table~\ref{tb:types}.
 
\vspace{3pt}

\begin{center}
\begin{figure}[t]
\includegraphics[angle=270,width=0.45\textwidth]{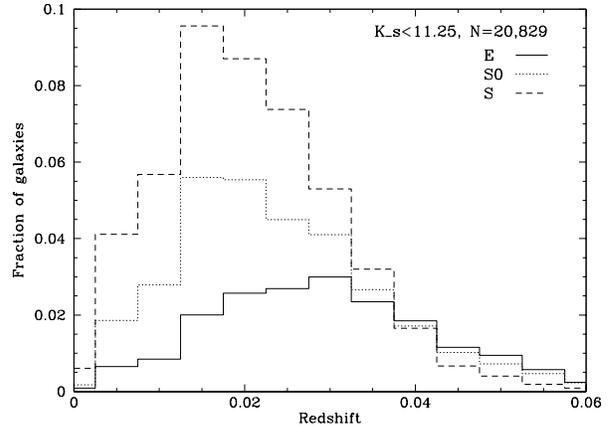}
\caption{Histogram of 2MRS11.25 galaxies as a function of redshift for the three main morphological classes.\\ \label{fig:histztype}}
\end{figure}
\end{center}

The distribution of the galaxies in 2MRS11.25 by morphological type is shown in Fig.~\ref{fig:histtype}, while Fig.~\ref{fig:histztype} shows histograms by redshift for the three broad morphological classes described above. While the histograms show the same pattern as Fig.~\ref{fig:zdist}, spirals dominate the dataset at lower redshifts, while ellipticals flatten near $z\approx 0.03$ and extend to higher redshifts, as expected given their higher luminosity.

\section{Previous results from 2MRS}

The 2MRS11.25 sample has been used in several publications:

\begin{itemize}
\item \citet{erdogdu06a} calculated the acceleration on the Local Group (LG). Their estimate of the dipole seems to converge to the cosmic microwave background (CMB) result within 60 h$^{-1}$~Mpc, suggesting that the bulk of the motion of the LG comes from structures within that distance. They also carried out an analysis of the dipole weighting the sample by its luminosity (rather than the counts) and found relatively minor changes. 
\vspace{6pt}
\item \citet{erdogdu06b} calculated density and velocity fields. All major local superclusters and voids were successfully identified, and backside infall on to the ``Great Attractor'' region (at 50~$h^{-1}$~Mpc) was detected.
\end{itemize}

\begin{itemize}
\vspace{-24pt}
\item \citet{westover07} measured the correlation function and found a steeper relationship between galaxy bias and luminosity than previously determined for optical samples, implying that near-infrared luminosities may be better mass tracers than optical ones. The relative biasing between early- and late-type galaxies was best fit by a power-law with no improvement when stochasticity was added, leaving open the possibility that populations of galaxies may evolve between one another.
\vspace{6pt}
\item \citet{crook07} produced a catalog of galaxy groups, which was later used to model the local velocity field in \citet{crook10}.
\vspace{6pt}
\item \citet{erdogdu09} predicted the acceleration of the Local Group generated by 2MRS in the framework of $\Lambda$CDM and the halo model of galaxies. Their analysis suggested that it is not necessary to invoke additional unknown mass concentrations to explain the misalignment between the CMB velocity vector and the 2MRS dipole.
\vspace{6pt}
\item \citet{lavaux10} derived the peculiar velocity field for 2MRS11.25 using an orbit-reconstruction algorithm and estimated the mean matter density within 3,000 km/s to be $\Omega_m=0.31\pm0.05$. They also studied the convergence toward the CMB dipole and found that less than half of the amplitude is generated within 40~$h^{-1}$~Mpc.

\item \citet{davis11} compared 2MRS11.25 to the SFI$++$ peculiar velocity survey \citep{masters06,springob07} to place constraints on the bias between galaxies and dark matter halos, as well as $\beta=f(\Omega)/b$ (where $f$ is the rate of growth of structure and $b$ is the bias factor) and $\sigma_8$ (which measures the amplitude of the linear power spectrum on the scale of 8~$h^{-1}$~Mpc).
\end{itemize}

\section{Summary}

2MASS has fulfilled its goal of providing and extremely uniform, deep and unbiased survey of the nearby Universe. The 2MASS Redshift Survey is 97.6\% complete to a limiting magnitude of K$_s=11.75$~mag over 91\% of the sky, and its catalog contains redshifts for 43,533 galaxies.

\vspace{6pt}

2MRS has produced an essentially complete map of the local Universe out to $z\sim0.08$. While the characteristics of the structures are similar to what has been seen before, we now have a nearly full view of the nearby Universe. Now we need to measure not only the redshifts, but also real distances \citep[e.g.,][]{masters08} to extract the full measure of cosmological information.  
  
\begin{acknowledgements}
This paper was written in part while JPH was a Sackler visitor at the Institute of Astronomy, Cambridge, UK.  Thanks are also due to the staff at the Fred.L.~Whipple, Cerro Tololo and McDonald Observatories, and the entire 2MASS team. 

\vspace{3pt}

JPH, KLM and ACC acknowledge support by the National Science Foundation under grant AST-0406906 and by the Smithsonian Institution.

\vspace{3pt}

LMM acknowledges support by the Smithsonian Institution Visiting Scholar program, by NASA through Hubble Fellowship grant HST-HF-01153 from the Space Telescope Science Institute, by the National Science Foundation through a Goldberg Fellowship from the National Optical Astronomy Observatory, and by the Texas A\&M University Mitchell-Heep-Munnerlyn Endowed Career Enhancement Professorship in Physics or Astronomy.

\vspace{3pt}

KLM acknowledges funding from the Leverhulme Trust as a 2010 Early Career Fellow and from the Peter and Patricia Gruber Foundation as the 2008 IAU Fellow.

\vspace{3pt}

CMH was supported in part by a National Science Foundation Research Experience for Undergraduates under Grant No. 9731923.

\vspace{3pt}

OL acknowledges support from a Royal Society Wolfson Research Merit Award.

\vspace{3pt}

JPH and LMM were visiting astronomers at Cerro Tololo Inter-American Observatory, operated by the Association of Universities for Research in Astronomy under contract with the National Science Foundation.

\vspace{6pt}

This publication has made use of the following resources:

\vspace{3pt}

\begin{itemize}

\item data products from the Two Micron All Sky Survey, which is a joint project of the University of Massachusetts and the Infrared Processing and Analysis Center at the California Institute of Technology, funded by the National Aeronautics and Space Administration and the National Science Foundation.

\item the NASA/IPAC Extragalactic Database (NED) which is operated by the Jet Propulsion Laboratory, California Institute of Technology, under contract with the National Aeronautics and Space Administration.

\item the 6dF Galaxy Survey (DR3), supported by Australian Research Council Discovery–Projects Grant (DP-0208876). The 6dFGS web site is \url{http://www.aao.gov.au/local/ www/6df/}.

\item the Sloan Digital Sky Survey III (DR8). Funding for SDSS-III has been provided by the Alfred P. Sloan Foundation, the Participating Institutions, the National Science Foundation, and the U.S. Department of Energy. The SDSS-III web site is \url{http://www.sdss3.org/}. SDSS-III is managed by the Astrophysical Research Consortium for the Participating Institutions of the SDSS-III Collaboration including the University of Arizona, the Brazilian Participation Group, Brookhaven National Laboratory, University of Cambridge, University of Florida, the French Participation Group, the German Participation Group, the Instituto de Astrofisica de Canarias, the Michigan State/Notre Dame/JINA Participation Group, Johns Hopkins University, Lawrence Berkeley National Laboratory, Max Planck Institute for Astrophysics, New Mexico State University, New York University, Ohio State University, Pennsylvania State University, University of Portsmouth, Princeton University, the Spanish Participation Group, University of Tokyo, University of Utah, Vanderbilt University, University of Virginia, University of Washington, and Yale University.

\item the VizieR catalog access tool operated at the CDS, Strasbourg, France.

\item the Digitized Sky Surveys, produced at the Space Telescope Science Institute under U.S. Government grant NAG W-2166. The images of these surveys are based on photographic data obtained using the Oschin Schmidt Telescope on Palomar Mountain and the UK Schmidt Telescope. 

\item NASA's Astrophysics Data System at the Harvard-Smithsonian Center for Astrophysics.

\end{itemize}

\vspace{3pt}

 Typing services provided by Fang, Inc.

\end{acknowledgements}

\vspace{12pt}

\noindent{{\it Facilities:} \facility{FLWO:1.5m (FAST)}, \facility{CTIO:1.5m (RCSpec)}, \facility{Blanco (RCSPec)}, \facility{Struve (es2)}, \facility{HET (LRS)}}

\bibliographystyle{apj}
\bibliography{huchra_2mrs}

\clearpage

\begin{appendix}
In this Appendix, we present all-sky plots of the 2MRS data set in
equatorial coordinates as well as ancillary tables.

\renewcommand{\thefigure}{A\arabic{figure}}
\begin{figure*}[h]
\begin{center}
\includegraphics[height=3.5in]{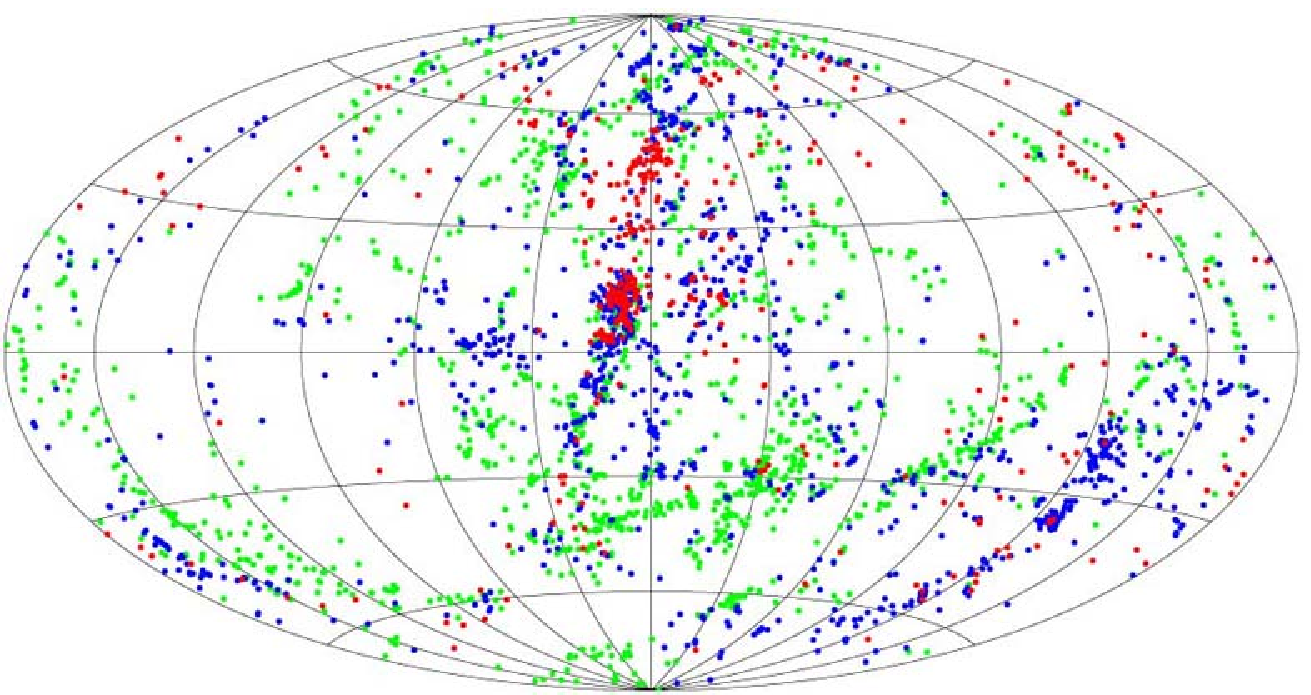}%{huchra_fig6.ps}
\end{center}
\caption{2MASS galaxies inside the 3,000 km/s sphere in equatorial coordinates (centered at R.A.$=0^\circ$ and following the convention of R.A. increasing to the left). Heliocentric velocities are color coded with red, blue \& green representing bins of increasing redshift/distance.  Red for $V_h < 1,000$~km/s, blue for $1,000 < V_h < 2,000$~km/s, and green for $2,000 < V_h < 3,000$~km/s. \label{fig:A1}} 
\end{figure*}

\begin{figure*}
\begin{center}
\includegraphics[height=3.5in]{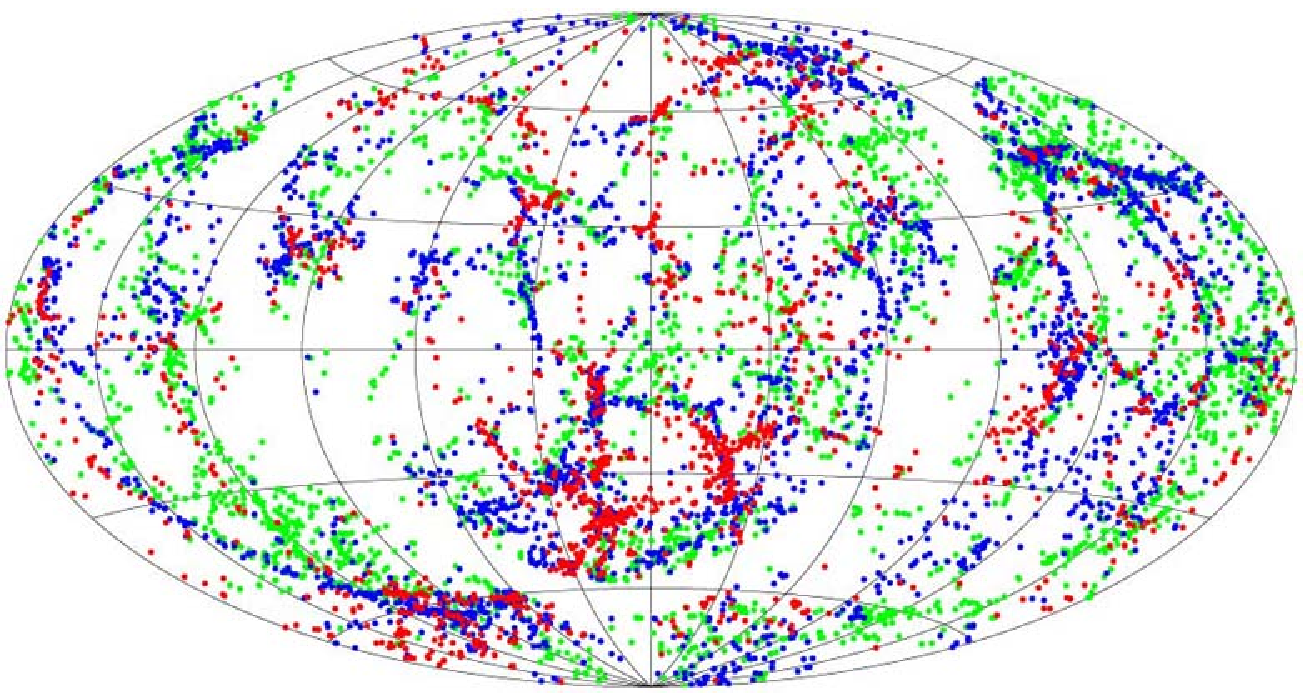}
\end{center}
\caption{Same as Fig.~\ref{fig:A1}, but for velocities between 3,000 and 6,000 km/s. \label{fig:A2}}
\end{figure*}

\clearpage

\vspace*{0.5in}

\begin{figure*}[h]
\begin{center}
\includegraphics[height=3.5in]{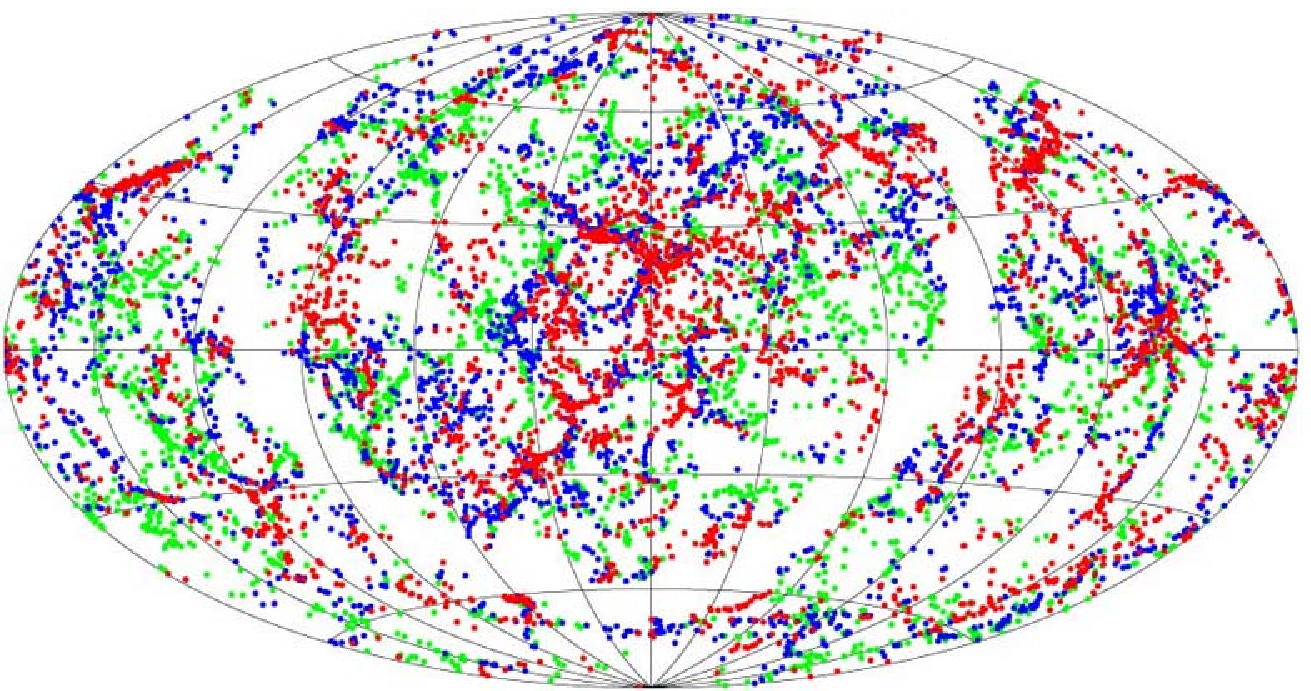}
\end{center}
\caption{Same as Fig.~\ref{fig:A2}, but for velocities between 6,000 and 9,000 km/s. \label{fig:A3}}
\end{figure*}

\vfill

\begin{figure*}[b]
\begin{center}
\includegraphics[height=3.5in]{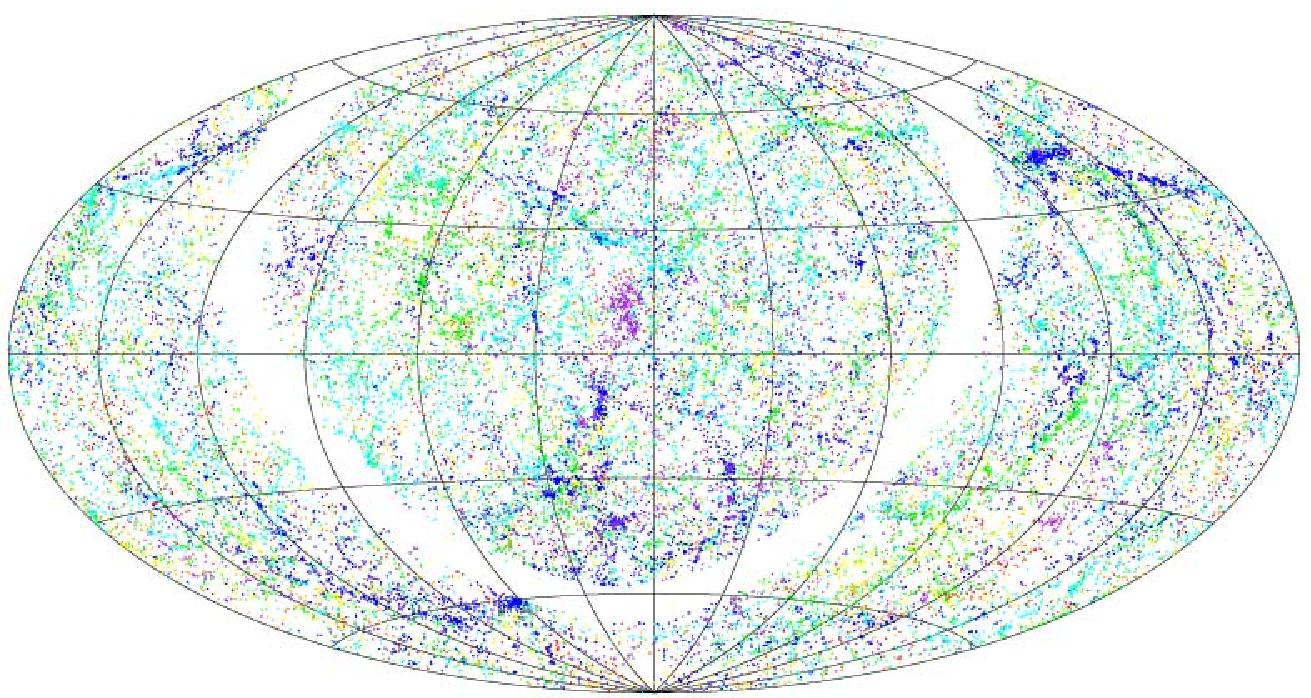}
\end{center}
\caption{Same as Fig.~\ref{fig:os4}, but in equatorial co-ordinates (centered at R.A.$=0^\circ$ and following the convention of R.A.~increasing to the left). \label{fig:A4}}
\end{figure*}
\clearpage
\renewcommand{\thetable}{A\arabic{table}}
\begin{deluxetable*}{ll}[h]
\tablewidth{0pc}
\tablecaption{2MASS XSC or LGA objects\\removed from input catalog \label{tb:rej}}
\tablehead{\colhead{2MASS ID} &\colhead{Reason for rejection}}
\startdata 
$00031127-5444588$ & Piece of galaxy 00031064-5444562\\
$00240535-7204531$ & Globular cluster in SMC\\
$00255209-0939420$ & Piece of galaxy 00255246-0939427\\
$00265282-7132113$ & Globular cluster in SMC\\
$00364578+2134078$ & Piece of galaxy 00364500+2133594\\
$00460635-0143434$ & Piece of galaxy 00460539-0143242\\
$00520075+6821243$ & Image flaw\\
$00523957-2637338$ & Star cluster (NGC 288)\\
$00524844-2637078$ & Star cluster (NGC 288)\\
$00525061-2635148$ & Star cluster (NGC 288)\\
$00525389-2635418$ & Star cluster (NGC 288)\\
$00584209+5628334$ & Image flaw\\
$01024864-0624482$ & Piece of galaxy 01024825-0624419\\
$01081982-7252599$ & Star cluster in SMC (NGC 419)\\
$01240782-7309037$ & HII region in SMC \\
$01244565+0959356$ & Piece of galaxy 01244588+0959403 \\
$01244585+0959406$ & Piece of galaxy 01244588+0959403 \\
$01500012+5354469$ & Image flaw \\
$01533610+4357488$ & Piece of galaxy 01533629+4357558 \\
$01550198-2600581$ & Piece of galaxy 01545983-2601101 \\
$02041855+2839171$ & Piece of galaxy 02041854+2839224 \\
$02115327+7026364$ & Piece of galaxy 02115039+7026384 \\
$02175760+1433128$ & Piece of galaxy 02175963+1432387 \\
$02213358+7038115$ & Piece of galaxy 02213090+7038118 \\
$02224353+4225299$ & Piece of galaxy 02223290+4220539
\enddata
\tablecomments{This table is presented in its entirety in the online version of the paper.}
\end{deluxetable*}

\begin{deluxetable*}{lrrrlrrr}
\tabletypesize{\scriptsize}
\tablewidth{0pc} 
\tablecaption{2MASS XSC objects with reprocessed photometry\label{tb:rep}}
\tablehead{\multicolumn{4}{c}{Original} & \multicolumn{4}{c}{Reprocessed}\\
\colhead{2MASS ID} & \colhead{$K^0_s$} & \colhead{$r_{\rm{iso}}$} & \colhead{$b/a$} & \colhead{2MASS ID} & \colhead{$K^0_s$} & \colhead{$r_{\rm{iso}}$} & \colhead{$b/a$}\\
\colhead{} & \colhead{(mag)} & \colhead{($\log_{10} \arcsec$)} & \colhead{} & \colhead{} & \colhead{(mag)} & \colhead{($\log_{10} \arcsec$)} & \colhead{}}
\startdata 
$00143065-0710028$ & 10.150 & 1.524 & 0.880 & $00143062-0710026$ &  9.976 & 1.529 & 0.810\\
$00144455-0720423$ &  8.771 & 1.721 & 0.700 & $00144460-0720430$ &  8.758 & 1.721 & 0.680\\
$00424581-2333406$ &  9.956 & 1.394 & 0.740 & $00424586-2333418$ &  9.959 & 1.338 & 0.780\\
$00510187-0703247$ &  9.317 & 1.531 & 0.400 & $00510186-0703249$ &  9.401 & 1.486 & 1.000\\
$00545028+2914482$ & 10.405 & 1.332 & 0.560 & $00545031+2914474$ & 10.392 & 1.318 & 0.570\\
$00564266-0954500$ &  9.588 & 1.695 & 0.860 & $00564267-0954507$ &  9.620 & 1.659 & 0.810\\
$01025144-6536359$ &  9.107 & 2.040 & 0.140 & $01025152-6536366$ &  9.145 & 1.989 & 0.154\\
$01243377+0143532$ &  8.801 & 1.818 & 0.980 & $01243380+0143522$ &  8.772 & 1.756 & 0.950\\
\enddata
\tablecomments{This table is presented in its entirety in the online version of the paper. All properties of the reprocessed galaxies are listed in Table~\ref{tb:cat} under their respective 2MASS IDs. The contents of this Table are intended to provide an overview of the changes due to the reprocessed photometry.}
\end{deluxetable*}

\clearpage

\begin{deluxetable*}{l}
\tablewidth{5in} 
\tablecaption{2MASS XSC galaxies with suspect photometry\\flagged for reprocessing at a later date\label{tb:flg}}
\tablehead{\colhead{2MASS ID}}
\startdata
$00364500+2133594$ \\
$01310193+4903364$ \\
$01595247-0705233$ \\
$02353199-0709366$ \\
$02352772-0921216$ \\
$02542739+4134467$ \\
$02594211+3602171$ \\
$03050558+4403167$ \\
$03235400-3730449$ \\
$03252491-1613594$ \\
$03261474+3803504$ \\
$03284660+3633226$ \\
$03392830+1323417$ \\
$03504306+7407352$\\
$03582336-4428024$\\
$04055073+3043032$\\
$04222904-5620157$\\
$04531024+3423105$\\
$05100496-1456598$\\
$05453957+5836099$\\
$05495771+5023330$\\
$06100970-3406499$\\
$06413093+4010268$\\
$07114179+4952003$\\
$07351844-5014596$
\enddata
\tablecomments{This table is presented in its entirety in the online version of the paper. All properties of the flagged galaxies are listed in Table~\ref{tb:cat} under their respective 2MASS IDs. This Table is only intended to provide an index of the galaxies with suspect photometry.}
\end{deluxetable*}

\begin{deluxetable*}{l}[h]
\tabletypesize{\scriptsize}
\tablewidth{5in} 
\tablecaption{2MASS XSC galaxies with compromised photometry\\flagged for reprocessing and removed from catalog\label{tb:flr}}
\tablehead{\colhead{2MASS ID}}
\startdata 
$00093966+5301008$\\
$00112204+0623372$\\
$00151096-2352551$\\
$00203715+2839334$\\
$00244982-4021070$\\
$00271148+5103476$\\
$00293249+5150575$\\
$00350737+4929082$\\
$00385467+0703458$\\
$00452378+5353203$\\
$00580474-8140329$\\
$01074777+7312247$\\
$01115336+2743193$\\
$01422276-1317152$\\
$01551704+5507250$\\
$01564822+3559412$\\
$02021375-6926541$\\
$02113146+5112023$\\
$02142526+4934260$\\
$02210796-0259054$\\
$02213090+7038118$\\
$02343012+1935176$\\
$02575308+1301508$\\
$03154885+4820359$\\
$03175766-4414175$\\
$03372681+0336579$
\enddata
\tablecomments{This table is presented in its entirety in the online version of the paper. These galaxies are not part of our catalog and therefore we do not list their photometric properties. This Table is only intended to provide other users of the 2MASS XSC an index of galaxies that we consider to have compromised photometry.}
\end{deluxetable*}

\clearpage
\begin{deluxetable*}{lrrclrrccl}
\tabletypesize{\scriptsize}
\tablewidth{6in}
\setlength{\tabcolsep}{0.04in}
\tablecaption{Alternative redshifts chosen over default NED redshifts (columns 1-10) \label{tb:nedz}}
\tablehead{
\colhead{(1)}      & \colhead{(2)}   & \colhead{(3)}  & \colhead{(4)}   & \colhead{(5)}  & \colhead{(6)}   & \colhead{(7)}  & \colhead{(8)}   & \colhead{(9)}    & \colhead{(10)}\\
\colhead{2MASS ID} & \multicolumn{4}{c}{NED} & \multicolumn{5}{c}{Alternative} \\
\colhead{} & \colhead{$v$} & \colhead{$\sigma_v$} & \colhead{Qual} & \colhead{Bibcode} & \colhead{$v$} & \colhead{$\sigma_v$} & \colhead{cat} & \colhead {sep}       & \colhead{Bibcode}\\
\colhead{} & \multicolumn{2}{c}{(km/s)}           & \colhead{}     & \colhead{}        & \multicolumn{2}{c}{(km/s)}           & \colhead{}    & \colhead {$\arcmin$} & \colhead{}}
\startdata 
\multicolumn{10}{l}{No redshift in NED for 2MASS XSC ID, but redshift exists under another ID}\\
\hline
$00014401-3025082$ &  \nd  & \nd &  \nd &        \nd          &  8604 &  64 & M & 0.0 & 20032dF...C...0000C \\
$00062997-3218179$ &  \nd  & \nd &  \nd &        \nd          & 13531 & 150 & M & 0.0 & 1998MNRAS.300..417R \\
$00072080-2807072$ &  \nd  & \nd &  \nd &        \nd          & 18048 &  54 & M & 0.2 & 1998AJ....116....1D \\
$00084652-3100399$ &  \nd  & \nd &  \nd &        \nd          & 16609 &  89 & M & 0.0 & 20032dF...C...0000C \\
$00105717-3512292$ &  \nd  & \nd &  \nd &        \nd          & 14910 &  30 & M & 0.0 & 1986MNRAS.220..901P \\
$00135592+0342508$ &  \nd  & \nd &  \nd &        \nd          & 11802 &  32 & M & 0.0 & 1999ApJS..121..287H \\
$00150718-2912377$ &  \nd  & \nd &  \nd &        \nd          &  7675 &  64 & M & 0.0 & 20032dF...C...0000C \\
\enddata
\tablecomments{This table is presented in its entirety in the online version of the paper. A few representative lines are provided here for guidance on its format and contents. Code for column 8: NED position [M]atch, [O]ther sources in ZCAT.}
\end{deluxetable*}

\clearpage

\addtocounter{table}{-1}
\begin{deluxetable*}{ll}
\tabletypesize{\scriptsize}
\tablewidth{0pc}
\tablecaption{Alternative redshifts chosen over default NED redshifts (column 11)}
\tablehead{\colhead{(1)}      & \colhead{(11)}\\
\colhead{2MASS ID} &\colhead{Comments}\\
\colhead{} & \colhead{}\\
\colhead{} & \colhead{}}
\startdata 
\multicolumn{2}{l}{No redshift in NED for 2MASS XSC ID, but redshift exists under another ID}\\
\hline
$00014401-3025082$ & \\
$00062997-3218179$ & \\
$00072080-2807072$ & \\
$00084652-3100399$ & \\
$00105717-3512292$ & \\
$00135592+0342508$ & \\
$00141513-2816502$ & \\
\enddata
\tablecomments{This table is presented in its entirety in the online version of the paper. A few representative lines are provided here for guidance on its format and contents.}
\end{deluxetable*}

\clearpage

\begin{deluxetable*}{lrrrr}
\tablewidth{0pc} 
\tablecaption{Redshifts from 6dFGS, SDSS or NED\\for galaxies also observed by 2MRS\label{tb:altz}}
\tablehead{\colhead{2MASS ID} &\colhead{$v$} & \colhead{$\sigma(v)$} & \colhead{Vel} & \colhead{Bibcode}\\
\colhead{} & \colhead{(km/s)} & \colhead{(km/s)} & \colhead{src} & \colhead{}}
\startdata 
$00010597-5359303$ &  9415 &  45 & 6 & 20096dF...C...0000J\\
$00021610-2926230$ & 18334 &  45 & 6 & 20096dF...C...0000J \\
$00023474-2948240$ & 17753 &  45 & 6 & 20096dF...C...0000J \\
$00023794+1638377$ &  6350 &  19 & N & 1999PASP..111..438F\\
$00031064-5444562$ &  9790 &  45 & 6 & 20096dF...C...0000J\\
$00032067-3008493$ & 20304 &  45 & 6 & 20096dF...C...0000J \\
$00034964+0203594$ & 29320 & \ndr& N & 2000ApJS..129..547B\\
$00042463-5257316$ &  9851 &  45 & 6 & 20096dF...C...0000J\\
\enddata
\tablecomments{This table is presented in its entirety in the online version of the paper. Code for column 4: [N]ED 2MASS ID match, [S]DSS-DR8, [6]dFGS.}
\end{deluxetable*}

\clearpage

\begin{longtable}{ll}
\tablewidth{0pc} 
\tablecaption{2MRS Catalog -- bibliographic references\label{tb:bib}}
\tablehead{\colhead{ADS bibcode} & \colhead{Reference}}
1969MSAIt..40..559B & \citet{1969MSAIt..40..559B}\\
1970ApJ...160..405S & \citet{1970ApJ...160..405S}\\
1970ApJ...160L..33B & \citet{1970ApJ...160L..33B}\\
1971ApJ...168..321C & \citet{1971ApJ...168..321C}\\
1971CGPG..C...0000Z & \citet{1971cscg.book.....Z}\\
1972ApJ...172L..37B & \citet{1972ApJ...172L..37B}\\
1972ApJ...173..247S & \citet{1972ApJ...173..247S}\\
1972AuJPh..25..233W & \citet{1972AuJPh..25..233W}\\
\end{longtable}

\begin{deluxetable*}{cl}[b]
\tabletypesize{\scriptsize}
\tablewidth{0pc}
\tablecaption{Morphological Type Codes used in 2MRS \label{tb:types}}
\startdata 
\hline
\multicolumn{2}{l}{Types} \\
\hline
-9 & QSO/AGN                 \\
-7 & Unclassified Elliptical \\
-6 & Compact Elliptical      \\
-5 & E, and dwarf E          \\
-4 & E/SO                    \\
-3 & L-, SO-                 \\
-2 & L, SO                   \\
\enddata
\tablecomments{The morphological information is encoded in Table~\ref{tb:cat} following the ZCAT convention. It is a 5 digit code (I2, A1, I1, A1). The first two digits are the numerically coded T type, the next letter (if present) is the Bar type, the next digit (if present) is the numericaly coded luminosity class, and the final letter (if present) denotes morphological perculiarities.}
\end{deluxetable*}

\clearpage

\begin{deluxetable*}{lrrrrcl}
\tablewidth{0pc} 
\tablecaption{Redshifts for galaxies in the 2MASS XSC beyond the main 2MRS catalog limits\label{tb:zextra}}
\tablehead{
\colhead{2MASS ID} & \colhead{R.A.}  &\colhead{Dec.} & \colhead{$v$}   & \colhead{$\sigma(v)$} & \colhead{Vel} & \colhead{Bibliographic} \\
\colhead{}         & \multicolumn{2}{c}{(deg)}       & \multicolumn{2}{c}{(km/s)} & \colhead{src.} & \colhead{code}}
\startdata
$00000256+0817537$ &  0.01063 &   8.29817 & 11721 &  37 & O & 20112MRS.MMT..0000H\\
$00000896+0817338$ &  0.03729 &   8.29272 & 12280 &  36 & O & 20112MRS.MMT..0000H\\
$00001215+0205503$ &  0.05069 &   2.09740 &  6506 &  35 & O & 20112MRS.JPH..0000H\\
$00005299+0803392$ &  0.22079 &   8.06090 & 11917 &  35 & O & 20112MRS.JPH..0000H\\
$00005467+0803442$ &  0.22779 &   8.06231 & 11952 &  10 & O & 20112MRS.JPH..0000H\\
$00015848+1203580$ &  0.49370 &  12.06618 & 60938 &  55 & O & 20112MRS.JPH..0000H\\
$00021610-2926230$ &  0.56700 & -29.43970 & 18213 &  40 & C & 20112MRS.CTIO.0000H\\
$00023474-2948240$ &  0.64483 & -29.80667 & 17766 &  31 & C & 20112MRS.CTIO.0000H\\
\enddata
\tablecomments{This table is presented in its entirety in the online version of the paper. Codes for column 6: [C]TIO, Mc[D]onald, [F]LWO.}
\end{deluxetable*}

\clearpage

\begin{deluxetable*}{lrrrrcl}
\tablewidth{0pc} 
\tablecaption{Redshifts for galaxies not in the 2MASS XSC which were observed serendipitously\label{tb:znocat}}
\tablehead{
\colhead{2MASS ID$^\dagger$} & \colhead{R.A.}  &\colhead{Dec.} & \colhead{$v$}   & \colhead{$\sigma(v)$} & \colhead{Vel} & \colhead{Bibliographic} \\
\colhead{}         & \multicolumn{2}{c}{(deg)}       & \multicolumn{2}{c}{(km/s)} & \colhead{src.} & \colhead{code}}
\startdata
$13294821-2340551$ & 202.45088 & -23.68197 &  5029 & 52 & C & 20112MRS.CTIO.0000H\\
$10245589-1726331$ & 156.23289 & -17.44251 &  7811 & 43 & C & 20112MRS.CTIO.0000H\\
$10104329-1530289$ & 152.68038 & -15.50803 &  8388 & 29 & C & 20112MRS.CTIO.0000H\\
$01574403-0649389$ &  29.43346 &  -6.82746 &  8434 & 21 & C & 20112MRS.CTIO.0000H\\
$10324638-1238045$ & 158.19324 & -12.63459 &  8491 & 45 & C & 20112MRS.CTIO.0000H\\
$10324555-1238140$ & 158.18977 & -12.63721 &  8579 & 43 & C & 20112MRS.CTIO.0000H\\
$11111075-0628055$ & 167.79479 &  -6.46819 &  8676 & 14 & C & 20112MRS.CTIO.0000H\\
$03452702-0728392$ &  56.36260 &  -7.47756 & 10504 & 31 & C & 20112MRS.CTIO.0000H\\
$16265144-7709323$ & 246.71447 & -77.15896 & 13333 & 41 & C & 20112MRS.CTIO.0000H\\
$16311208-2615525$ & 247.80033 & -26.26458 & 13380 & 55 & C & 20112MRS.CTIO.0000H\\
$03190550-2140362$ &  49.77291 & -21.67673 & 15624 & 32 & C & 20112MRS.CTIO.0000H\\
$01135502-3659473$ &  18.47924 & -36.99647 & 15814 & 56 & C & 20112MRS.CTIO.0000H\\
$16385916-6421059$ & 249.74644 & -64.35169 & 16320 & 46 & C & 20112MRS.CTIO.0000H\\
$23342933-6921154$ & 353.62221 & -69.35433 & 25413 & 49 & C & 20112MRS.CTIO.0000H
\enddata
\tablecomments{$\dagger$: pseudo-2MASS ID generated from the celestial coordinates of the object.}
\end{deluxetable*}
\end{appendix}
\end{document}